\documentclass{jfm}
\pdfoutput=1
\usepackage[pdftex]{graphicx}
\usepackage{amsmath,amssymb,amsbsy,epsfig}
\usepackage[a4paper, left=2cm, right=2cm, top=2cm, bottom=2cm]{geometry}
\usepackage{natbib}

\def \bd {\boldsymbol}

\title[Numerics of Viscous Flows in Domains with Corners]{Viscous Flow in Domains with Corners:\\ Numerical Artifacts, their Origin and Removal}

\author[J.E. Sprittles and Y.D. Shikhmurzaev]{J\ls A\ls M\ls E\ls S\ns E.\ns S\ls P\ls R\ls I\ls
T\ls T\ls L\ls E\ls S\ns \and Y\ls U\ls L\ls I\ls I\ns D.\ns S\ls H\ls I\ls K\ls H\ls M\ls U\ls
R\ls Z\ls A\ls E\ls V\footnote{Email: Y.D.Shikhmurzaev@bham.ac.uk}}

\affiliation{School of Mathematics, University of Birmingham, Birmingham,  B15 2TT, UK.}

\begin{document}

\label{firstpage} \maketitle

\begin{abstract}

We show that an attempt to compute numerically a viscous flow in a domain with a piece-wise smooth
boundary by straightforwardly applying well-tested numerical algorithms (and numerical codes based
on their use, such as COMSOL Multiphysics) can lead to spurious multivaluedness and nonintegrable
singularities in the distribution of the fluid's pressure. The origin of this difficulty is that,
near a corner formed by smooth parts of the piece-wise smooth boundary, in addition to the solution
of the inhomogeneous problem, there is also an eigensolution. For obtuse corner angles this
eigensolution  (a) becomes dominant and (b) has a singular radial derivative of velocity at the
corner. A method is developed that uses the knowledge about the eigensolution to remove
multivaluedness and nonintegrability of the pressure. The method is first explained in the simple
case of a Stokes flow in a corner region and then generalised for the full-scale unsteady
Navier-Stokes flow in a domain with a free surface.

\end{abstract}

\section{Introduction}

In many problems of computational mechanics it is necessary to consider domains with piecewise
smooth boundaries, in particular the situation where two parts of the boundary locally form a
two-dimensional wedge. In elasticity such problems occur in fracture mechanics when one considers
how a crack propagates into a material \cite[see][]{moes99}.  A typical example from fluid dynamics
is found in the process of dynamic wetting \cite[see][]{dussan79,blake06}, in which a liquid
spreads over the surface of a solid, with the liquid-fluid free surface forming what is referred to
as a `contact angle' with the solid substrate. The mathematical problems in wedge regions that one
encounters in fluid and solid mechanics have much in common; below we shall consider some generic
computational issues in the context of fluid mechanics.

It has been known for a long time \cite[see][]{huh71} that classical boundary conditions, when
applied in dynamic wetting, i.e.\ in the situation where the `contact line' formed by the free
surface with the solid boundary moves with respect to the solid, lead to a physically unacceptable
outcome, namely a nonintegrably infinite force acting between the liquid and the solid. Termed the
`moving contact-line problem', this difficulty has become the subject of many theoretical works
(see Ch.~3 of \cite{shik07} for a review).

The common feature of almost all theoretical models proposed for the moving contact-line problem is
that the no-slip boundary condition on the solid surface, i.e.\ the condition for the tangential
component of the fluid's velocity, is replaced by a more general condition formulated for the
tangential stress (and hence allowing for slip). The impermeability condition for the normal
component of velocity on the solid surface as well as both the zero tangential-stress and
impermeability conditions on the free surface remain intact. As a result, one arrives at a
mathematical problem, locally in a wedge region, with different boundary conditions on the two
sides, both of which involve first-order differential operators applied to the bulk velocity
component tangential to the boundary and linear homogeneous algebraic conditions for the normal
component.

In the present work, we show that an attempt to directly apply well-tested standard numerical
methods to this problem can lead to a fundamentally unsatisfactory outcome. As is shown, even in
the simplest case of a steady Stokes flow in a corner region, irrespective of a particular
numerical implementation, in a certain range of wedge angles the computed bulk pressure is both
nonintegrably singular and multivalued as one approaches the corner. This outcome has a serious
`practical' implication: it becomes impossible to compute a numerical solution to the corresponding
free-surface problems with moving contact lines, as no capillary pressure would be able to balance
a nonintegrable bulk pressure. Alternatively, if the pressure singularity and multivaluedness are
suppressed using numerical means, the resulting `solution' becomes mesh-dependent and hence is by
no means a uniformly valid approximation of the solution of the original partial differential
equations. Neither of these options is satisfactory, and the persistent nature of the problem makes
it necessary to handle it robustly, beginning by finding out its origin.

The structure of the present paper is as follows. In Section~\ref{pf}, we formulate the problem to
be solved in the simplest case of a Stokes flow in a corner region and in Section~\ref{ana}
describe the local analytic asymptotic behaviour of the solution near the corner. In
Section~\ref{num}, numerical results are presented. They show that the numerical solution is in
excellent agreement with the analytic asymptotics for acute corner angles and completely incorrect
for obtuse angles. In particular, the fluid's pressure appears to be multivalued at the corner with
a nonintegrably singular behaviour near it. As demonstrated, the same outcome follows from a
commercially available piece of software, COMSOL Multiphysics. An analysis of the origin of the
problem, showing that the multivaluedness of pressure together with its nonintegrability is not
inherent in the mathematical formulation and has a spurious/numerical nature, is given in
Section~\ref{ori}. In Section~\ref{rem}, a method of removing this multivaluedness is formulated
and numerically verified, first, in the wedge geometry for the steady Stokes flow and then for
steady and unsteady Navier-Stokes equations in a full-scale dynamic wetting simulation. In
Section~\ref{sec:conclusion}, the results are summarized and some their implications are discussed.

\section{\label{pf}Problem formulation}

Consider the two-dimensional steady viscous flow of an incompressible Newtonian fluid, with density
$\rho$ and viscosity $\mu$, in a corner confined by straight boundaries located at $\theta=0$ and
$\theta=\alpha$ of a polar coordinate system $(r,\theta)$ in the plane of flow and the `far field'
boundary on an arc of a sufficiently large radius $r=R$. Conventionally, we will refer to the
$\theta=0$ and $\theta=\alpha$ boundaries as the `solid boundary' and the `free surface',
respectively. The `free surface' will be made genuinely free, with its shape to be determined, in
\S\ref{6.2}.

The flow is driven by the relative motion of a solid at $\theta=0$, which slides with speed $U$
parallel to itself, and, possibly, also by the far-field conditions. For simplicity, we will assume
that the velocity and length scales that characterize the flow are such that the Reynolds number
$\hbox{\it Re\/}$ based on these scales is small. Then as $\hbox{\it Re\/}\to0$, to leading order
in $\hbox{\it Re\/}$  we may consider the Stokes flow. It should be emphasized that all essential
results remain valid for the full Navier-Stokes equations since they come from the asymptotic
behaviour of the solution as $r\to0$. Considering the Stokes flow allows us to demonstrate the
method we use to handle the pressure multivaluedness more clearly, without additional but
nonessential details associated with handling nonlinear convective terms. These details are
described in \S\ref{6.2}.

For the problem in question, the non-dimensional Stokes equations for the bulk pressure $p$ and the
radial and azimuthal components of velocity $(u,v)$ take the form
\begin{equation}
\label{contin} \frac{1}{r}\frac{\partial(ru)}{\partial r}
+\frac{1}{r}\frac{\partial v}{\partial\theta}=0,
\qquad\qquad(0<r<R,\ 0<\theta<\alpha),
\end{equation}
\begin{subeqnarray}
\label{motion_prim}\gdef\thesubequation{\theequation \textit{a,b}} \frac{\partial p}{\partial
r}=\Delta u -\frac{u}{r^2}-\frac{2}{r^2}\frac{\partial v}{\partial\theta}, \qquad
\frac{1}{r}\frac{\partial p}{\partial\theta}=\Delta v -\frac{v}{r^2}+\frac{2}{r^2}\frac{\partial
u}{\partial\theta},
\end{subeqnarray}
where
\[
\Delta=\frac{\partial^2}{\partial r^2}+\frac{1}{r}\frac{\partial
}{\partial r} +\frac{1}{r^2}\frac{\partial^2}{\partial\theta^2}.
\]

On the solid surface, for a solution not to have multivalued velocity at the corner
\cite[see][]{dussan74,shik07}, we use the Navier slip condition \cite[see][]{navier23}, as opposed
to the no-slip condition of classical fluid mechanics, and keep intact the impermeability condition
for the normal component of velocity to the surface:
\begin{subeqnarray}
\label{vect_ss}\gdef\thesubequation{\theequation \textit{a,b}} \frac{\partial u}{\partial\theta}=
r\bar{\beta}(u-1),\quad v=0, \qquad\qquad(0<r<R,\ \theta=0).
\end{subeqnarray}
Here $\bar{\beta}$ is the dimensionless `coefficient of sliding friction' \cite[see][]{lamb32}. In
the limits $\bar{\beta}\to0$ and $\bar{\beta}\to\infty$, one recovers the conditions of zero
tangential stress (free slip) and no-slip, respectively. The value of $1/\bar{\beta}$ is
proportional to a (non-dimensional) `slip length' that characterizes the region where the velocity
field specified with the help of the Navier condition (\ref{vect_ss}\textit{a}) deviates from the
velocity field which would have been specified by no-slip.

On the free surface, we have the standard boundary conditions of
zero tangential stress and impermeability:
\begin{subeqnarray}
\label{vect_fs}\gdef\thesubequation{\theequation \textit{a,b}}  \frac{\partial
u}{\partial\theta}=0,\quad v=0,\qquad\qquad(0<r<R,\ \theta=\alpha).
\end{subeqnarray}

The boundary conditions in the far field can be imposed in different
ways. For simplicity, we will make the far-field conditions
`passive' and assume that:
\begin{equation}
\label{vect_far_field} \frac{\partial u}{\partial r}=\frac{\partial
v}{\partial r}=0, \qquad\qquad(r=R,\ 0<\theta<\alpha).
\end{equation}
This condition is an adaptation for a finite domain of a boundary condition that would specify the
asymptotic behaviour of the flow field at infinity. As is known \cite[see][]{moffatt64}, this
condition is satisfied if the Navier slip condition (\ref{vect_ss}\textit{a}) is replace by no-slip
(i.e.\ for $\bar{\beta}=\infty$ in (\ref{vect_ss}\textit{a})), so that (\ref{vect_far_field}) can
be seen as a condition that the disturbance caused by finiteness of $\bar{\beta}$ attenuates in the
far field. In computations, for the far-field to have a negligible effect on the near-field flow,
it is sufficient to put $R\ge100/\bar{\beta}$.

Equations (\ref{contin})--(\ref{vect_far_field}) fully specify the
problem of interest.

\section{\label{ana}Local asymptotics}

The defining feature of our problem is the angle formed by two parts of the boundary and, for
future references, it is useful to reproduce the leading-order asymptotics for the solution of
(\ref{contin})--(\ref{vect_far_field}) as $r\to0$ \cite[see][]{shik06,shik07}. After introducing
the stream function $\psi$ by
\begin{subeqnarray}
\label{streamfunction}\gdef\thesubequation{\theequation \textit{a,b}} u = \frac{1}{r}\frac{\partial
\psi}{\partial \theta}, \quad v = - \frac{\partial \psi}{\partial r}
\end{subeqnarray}
equations (\ref{contin})--(\ref{motion_prim}) are reduced to a
biharmonic equation $\Delta^2\psi=0$ with boundary conditions
(\ref{vect_ss})--(\ref{vect_fs}) taking the form
\begin{subeqnarray}\label{polar_navier_slip}
\gdef\thesubequation{\theequation \textit{a,b}} \frac{\partial^{2}\psi}{\partial\theta^{2}} =
r\bar{\beta}\left(\frac{\partial\psi}{\partial\theta}-r\right), \quad\psi=0,
\qquad\qquad(\theta=0,\ 0<r<R),
\end{subeqnarray}
\begin{subeqnarray} \label{polar_free_slip}
\gdef\thesubequation{\theequation \textit{a,b}} \frac{\partial^{2}\psi}{\partial\theta^{2}}=0,
\quad \psi=0, \qquad\qquad(\theta=\alpha,\ 0<r<R).
\end{subeqnarray}

Condition (\ref{polar_navier_slip}\textit{a}), which is the only inhomogeneous boundary condition
in the problem (i.e.\ the condition that drives the flow), suggests looking for the leading-order
term of the local asymptotics in the form $\psi=r^2F(\theta)$, which is one of a family of
separable solutions to the biharmonic equation of the form $\psi=r^\lambda F(\theta)$. After
substituting $\psi$ into the biharmonic equation and boundary conditions
(\ref{polar_navier_slip})--(\ref{polar_free_slip}), we have that
\begin{equation}
\label{stream_2} \psi = r^{2}
\left(B_{1}+B_{2}\theta+B_{3}\sin2\theta+B_{4}\cos2\theta\right),
\end{equation}
where the constants of integration $B_i~(i=1,\dots,4)$ are:
$B_1=-B_4=-\bar{\beta}/4$, $B_2=-B_1/\alpha$,
$B_3=B_1\cot2\alpha$.

The pressure field obtained from (\ref{motion_prim}) using
(\ref{stream_2}) and (\ref{streamfunction}) has the form
\begin{equation}\label{stream_2_pressure}
p=\frac{\bar{\beta}}{\alpha}\ln r + p_{0}.
\end{equation}
where $p_{0}$ is a constant which sets the pressure level.

The most notable characteristics of the flow are that, as $r\to0$,
the velocity scales {\it linearly\/} with $r$ whilst the pressure
is {\it logarithmically\/} singular at the corner and does not
depend on the angular coordinate $\theta$. These simple features
provide a quick test to determine whether there are any major
issues with a numerical scheme.

\section{\label{num}Numerical results}

The problem formulated in the previous section was incorporated into a numerical platform based on
the Galerkin finite element method. This platform, developed by the authors to tackle a range of
microfluidic capillary flows, has already been shown to provide novel results in the flow of
liquids over chemically patterned surfaces \cite[see][]{sprittles07,sprittles09}. As an additional
test of the robustness of our numerical results (and ubiquity of emerging numerical artifacts) for
the flows in the corner regions that we are examining here, these results have been verified for
test cases using a commercially available code, COMSOL Multiphysics.  All computations presented
below correspond to $\bar{\beta}=10$, $R=10$; the runs in the process of investigation covered a
wide range of parameters to ensure that the features described below are invariant with respect to
variations of these parameters.

\subsection{Details of computations}

The numerical solution of the steady Stokes problem (\ref{contin})--(\ref{vect_far_field}) uses
only a small degree of the platform's capability. Given the domain's geometry,  it was natural to
present the problem formulation in polar coordinates; this also made it easy to describe the
asymptotic results. However, for a generic free surface flow there is nothing to be gained from
this implementation and the finite element platform is formulated in the usual Cartesian
coordinates, with the domain's geometry utilised in the mesh generation (see below). As is
relatively standard in the finite element approximation of dynamic wetting flows
\cite[see][]{wilson06,lukyanov07}, we use the P2P1 triangular elements which approximate the
velocity quadratically and the pressure linearly, thus satisfying the LBB constraint
\cite[see][]{babuska72}.  A converged solution to the resulting linear equations is obtained after
one iteration, using for the inversion the MA41 solver provided by the Harwell Subroutine Library.

The skeleton of the finite element mesh is composed of circular arcs with the corner as their
centre, and radial-rays which run from $r=0$ to $r=R$. The elements are then tesselated around the
skeleton so as to create a structured mesh. Moving away from the corner, the distance between arcs
is gradually increased by $5\%$ so as to provide a high level of resolution near the corner whilst
ensuring that the problem remains computationally tractable. Due to the nature of the flow it has
been found that, as noted in previous investigations of a similar nature
\cite[see][]{wilson06,lukyanov08}, a large number of elements are required in the angular direction
near the corner in order to accurately capture the dynamics.

It can be seen in equation (\ref{stream_2_pressure}) that the pressure is logarithmically singular
as the corner is approached and, strictly speaking, one should look to incorporate special
`singular elements' there to capture this behaviour \cite[see][]{wilson06}. The platform we have
developed has the option of incorporating these singular elements but, in the results presented
here, we have opted against using them in order to simplify our exposition.  Their implementation
has no effect on the conclusions of the forthcoming sections, as verified by our test runs.

\subsection{Acute wedge angles}

First, we consider the case of $\alpha<\pi/2$ and show that our numerical results are in excellent
agreement with the local asymptotics described earlier.

\begin{figure}
\centering
\includegraphics*[viewport=0 50 350 300,angle=0,scale=0.7]{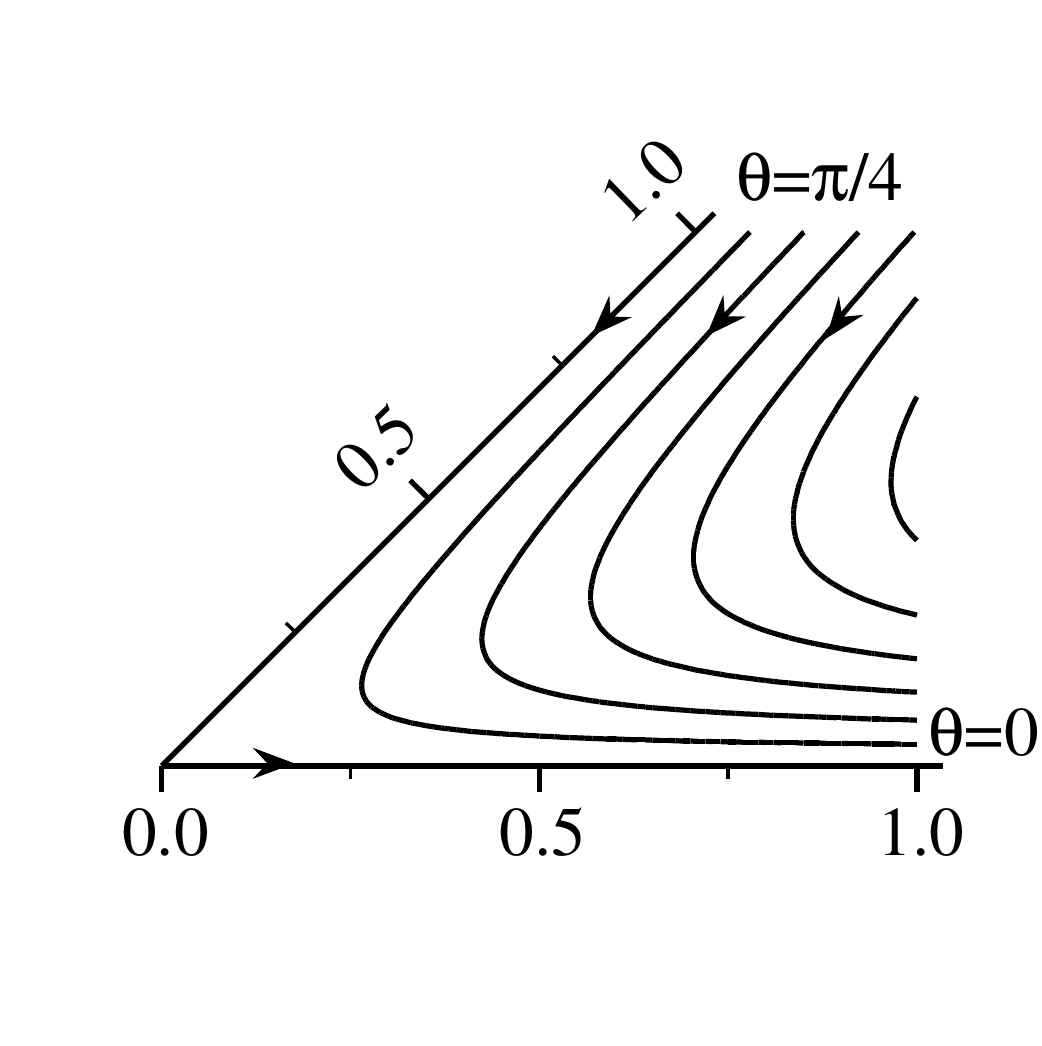} \caption{Streamlines for $\alpha=\pi/4$
in increments of $\psi=0.02$ with the boundaries at $\theta=0$ and at $\theta=\alpha$ corresponding
to $\psi=0$.} \label{F:45stream}
\end{figure}

The streamlines  in Fig.~\ref{F:45stream} illustrate the general features of the flow: motion is
created by the relative movement of the solid surface with respect to the corner. The fluid near
the solid is pulled out of the corner by the moving solid surface and, by continuity, it is
replenished  there due to the inflow from the far field.

\begin{figure}
\includegraphics*[viewport= -10 30 350 250  ,angle=0,scale=0.55]{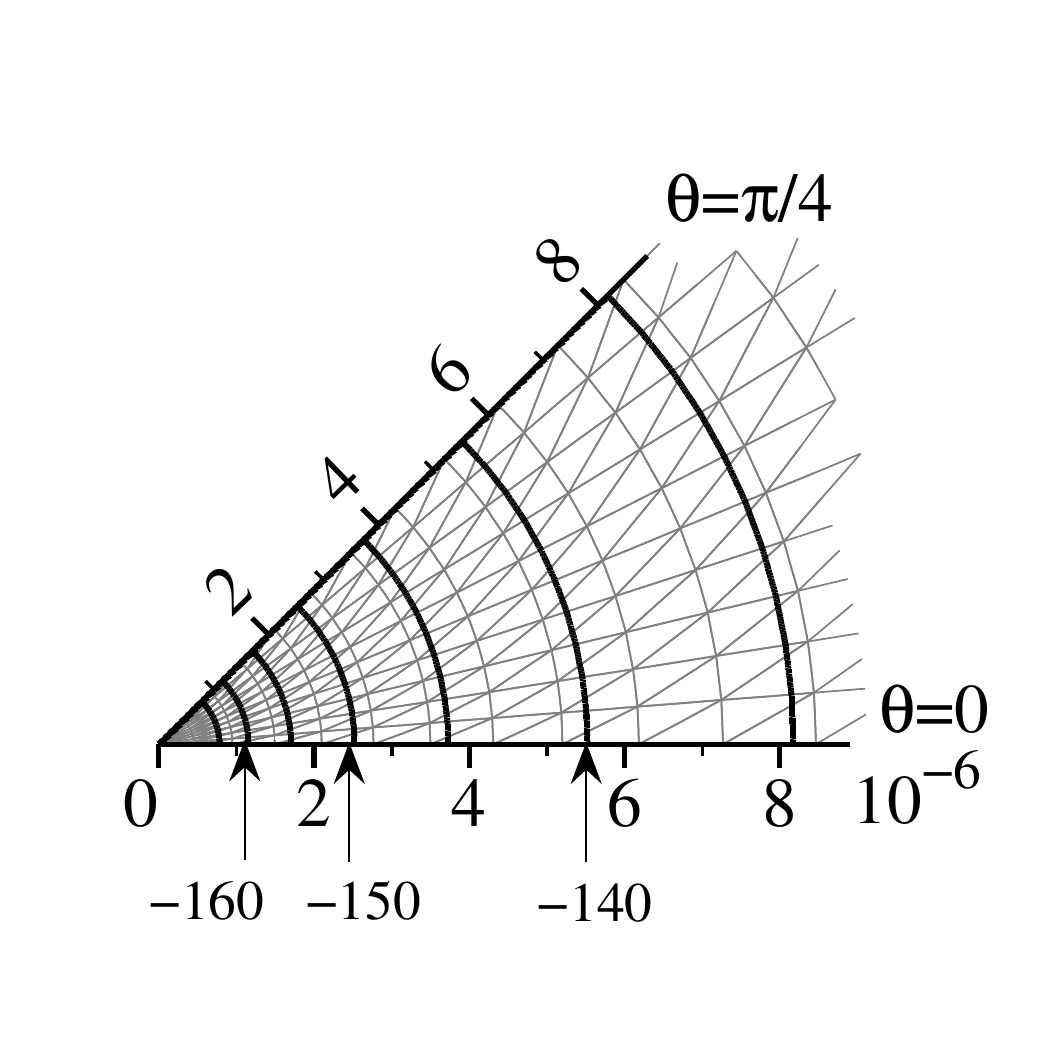}
\includegraphics*[viewport=-50 -10 550 250,angle=0,scale=0.55]{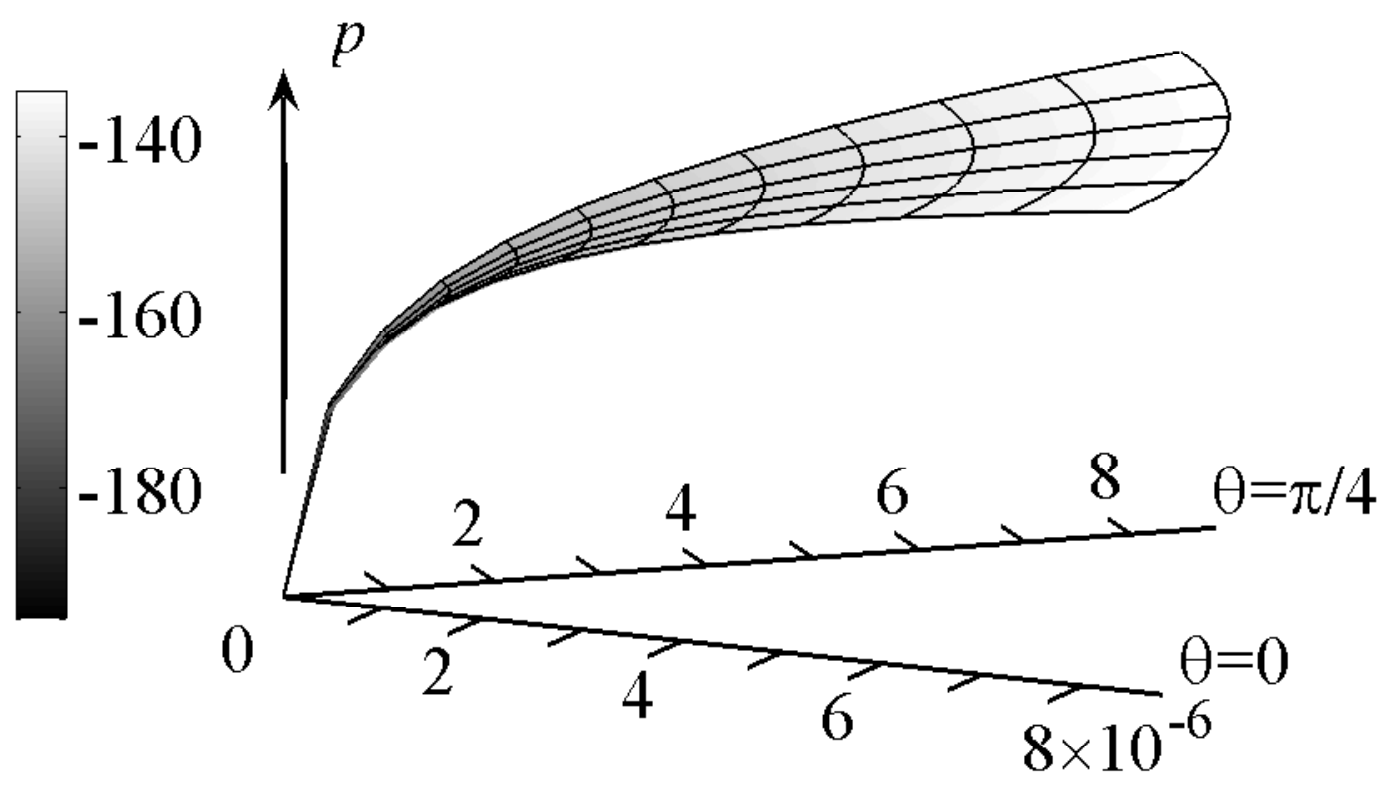}
\caption{Pressure distributions in the vicinity of the corner for $\alpha=\pi/4$.  Left: pressure
contours in steps of size 5, as the corner is approached with the underlying finite element mesh
visible. Right: surface plot of pressure.}\label{F:45pressure}
\end{figure}

The pressure distribution near the corner in the form of isobars and, to make it easier to
envisage, a 3-dimensional plot are shown in Fig.~\ref{F:45pressure}. As predicted by equation
(\ref{stream_2_pressure}) of the local asymptotics, the pressure in the vicinity of the corner is
independent of $\theta$.

\begin{figure}
\includegraphics*[viewport=-20 270 700 620,angle=0,scale=0.35]{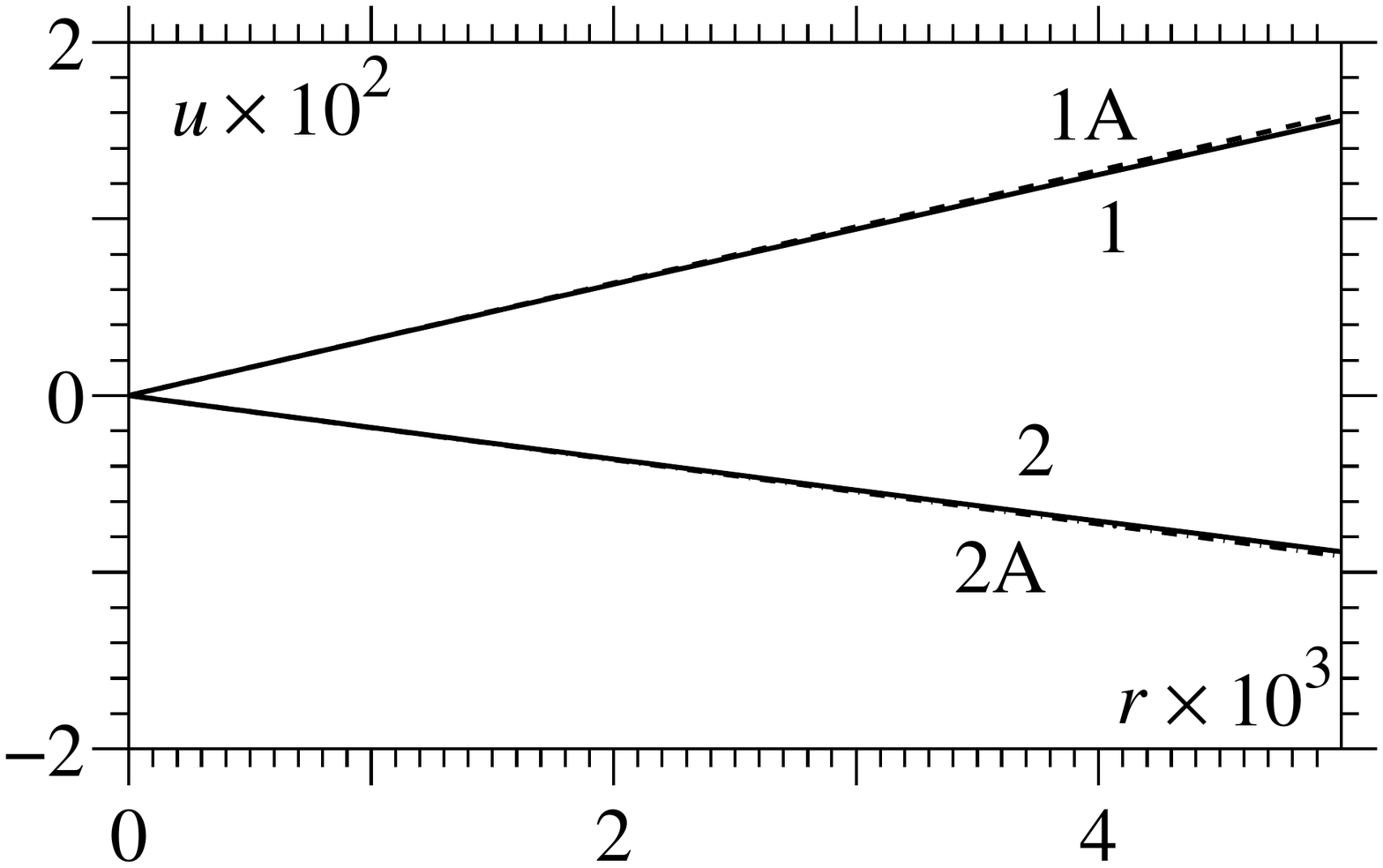}
\includegraphics*[viewport=-20 0 700 500,angle=0,scale=0.35]{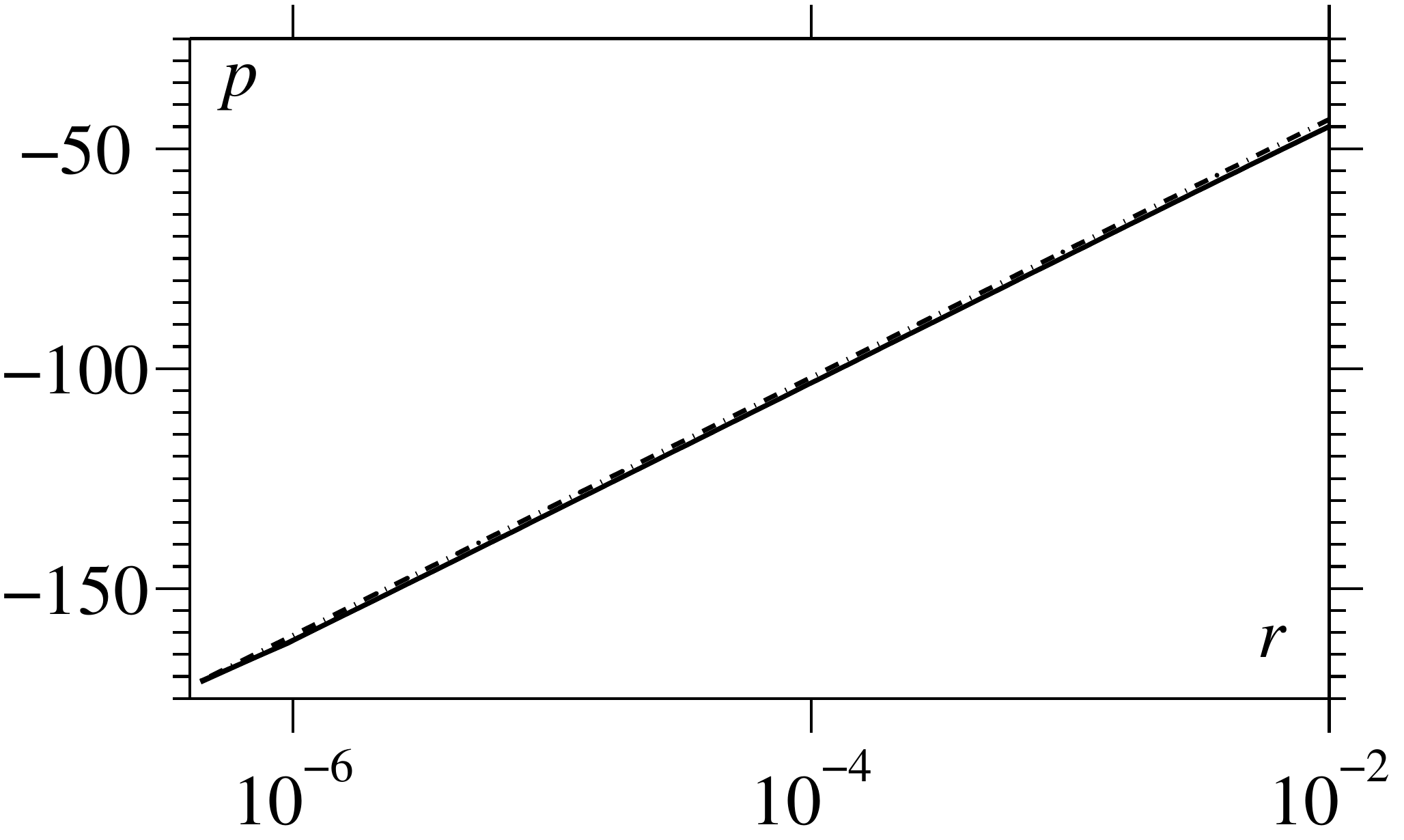}
\caption{Left: comparison of the computed radial velocity $u$ along the liquid-solid interface $1$
and liquid-gas interface $2$ with the corresponding asymptotic results $1A$ and $2A$, respectively.
Right: comparison of the computed pressure along the interfaces to the asymptotic result (dashed
line).}\label{F:45asymptotics}
\end{figure}

The quantitative comparison of the computed velocity and pressure with those given by the
asymptotics is shown in Fig.~\ref{F:45asymptotics}. As one can see, the agreement between numerical
and analytic results for the distribution of velocity and pressure is excellent. The velocity is
linear whilst the pressure, which is plotted in a semi-logarithmic frame, is logarithmic with the
expected gradient. The pressure constant $p_{0}$ in (\ref{stream_2_pressure}) has been chosen to
provide the best fit, but does not in any way determine the shape or gradient of the curve.

\subsection{Obtuse angles: multivaluedness of pressure}\label{eme}

In the previous subsection, we have shown that our algorithm gives excellent results for
$\alpha<\pi/2$. However, for the angles $\alpha>\pi/2$ the situation changes. The same code, as
well as COMSOL Multiphysics, that we used for comparison, produce results that are markedly
mesh-dependent, i.e.\ the numerical solution cannot be regarded as a uniformly valid approximation
of the analytic one.

\begin{figure}
\centering
\includegraphics*[viewport=100 40 600 300,angle=0,scale=0.7]{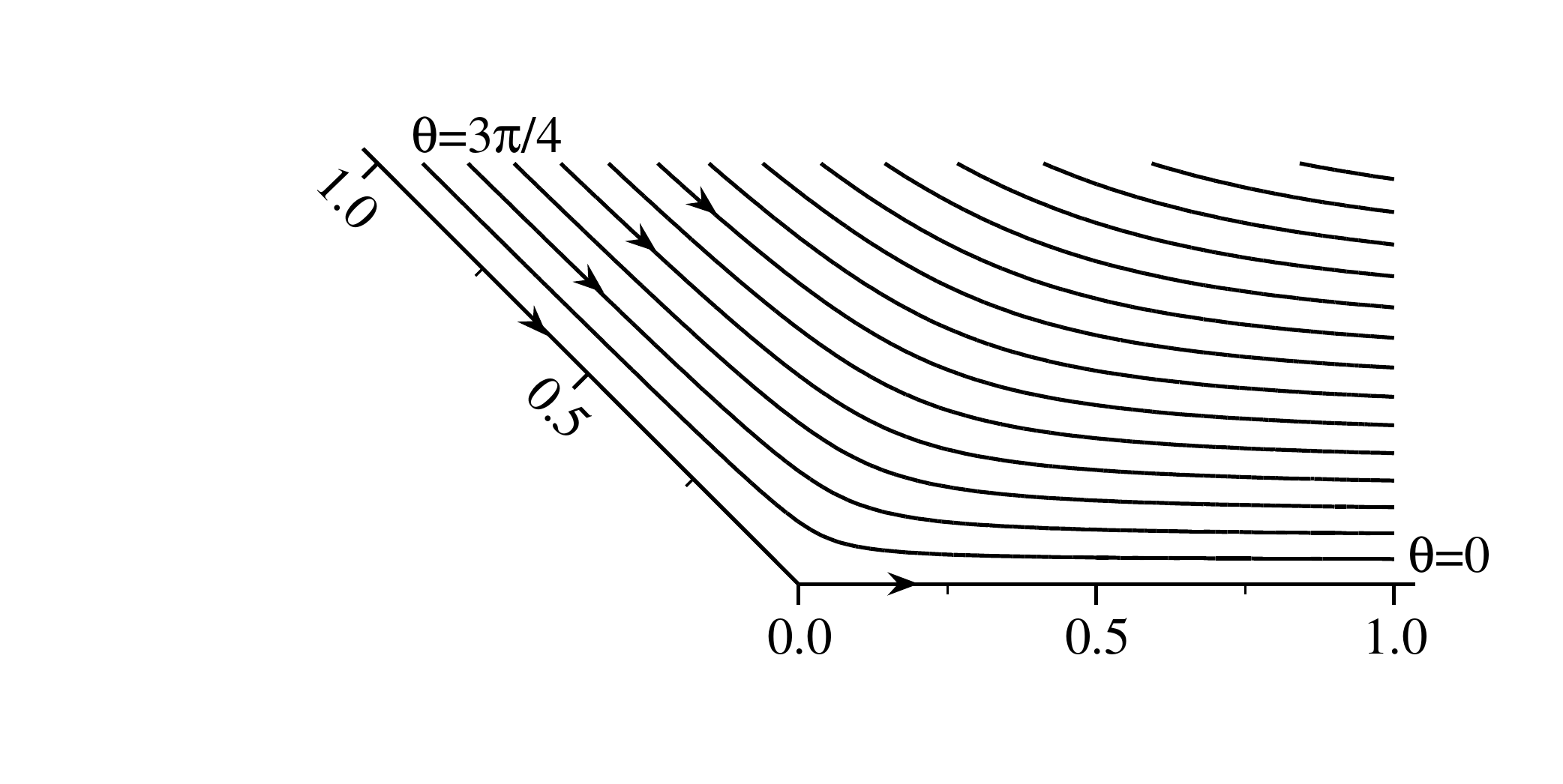}
\caption{Streamlines for $\alpha=3\pi/4$ in increments
of $\psi=0.04$ with the boundaries at $\theta=0$ and at $\theta=\alpha$ corresponding to $\psi=0$.}
\label{F:135eigstreamlines}
\end{figure}

\begin{figure}
\includegraphics*[viewport=100 -10 550 250,angle=0,scale=0.5]{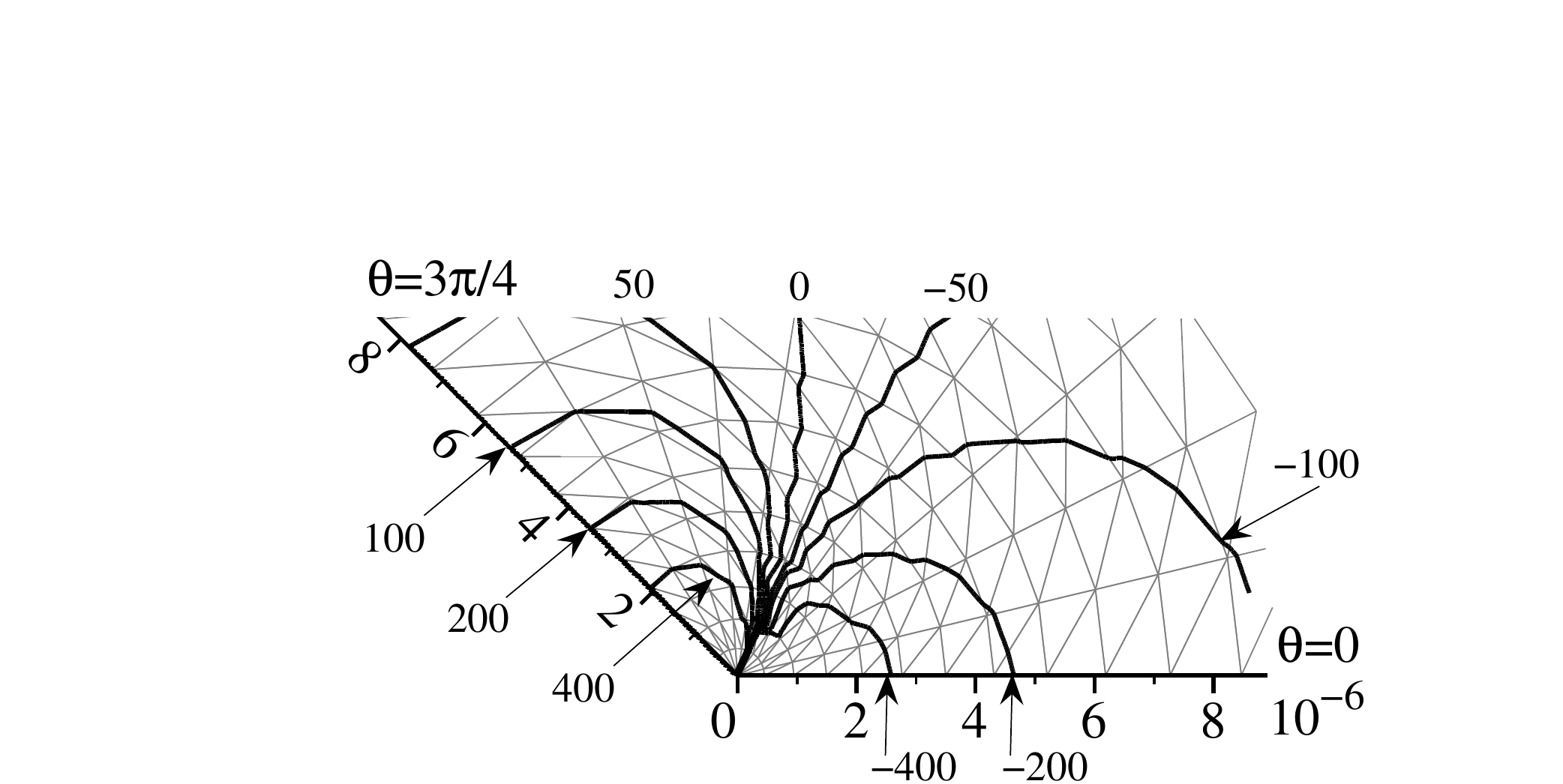}
\includegraphics*[viewport=-10 -20 550 280,angle=0,scale=0.5]{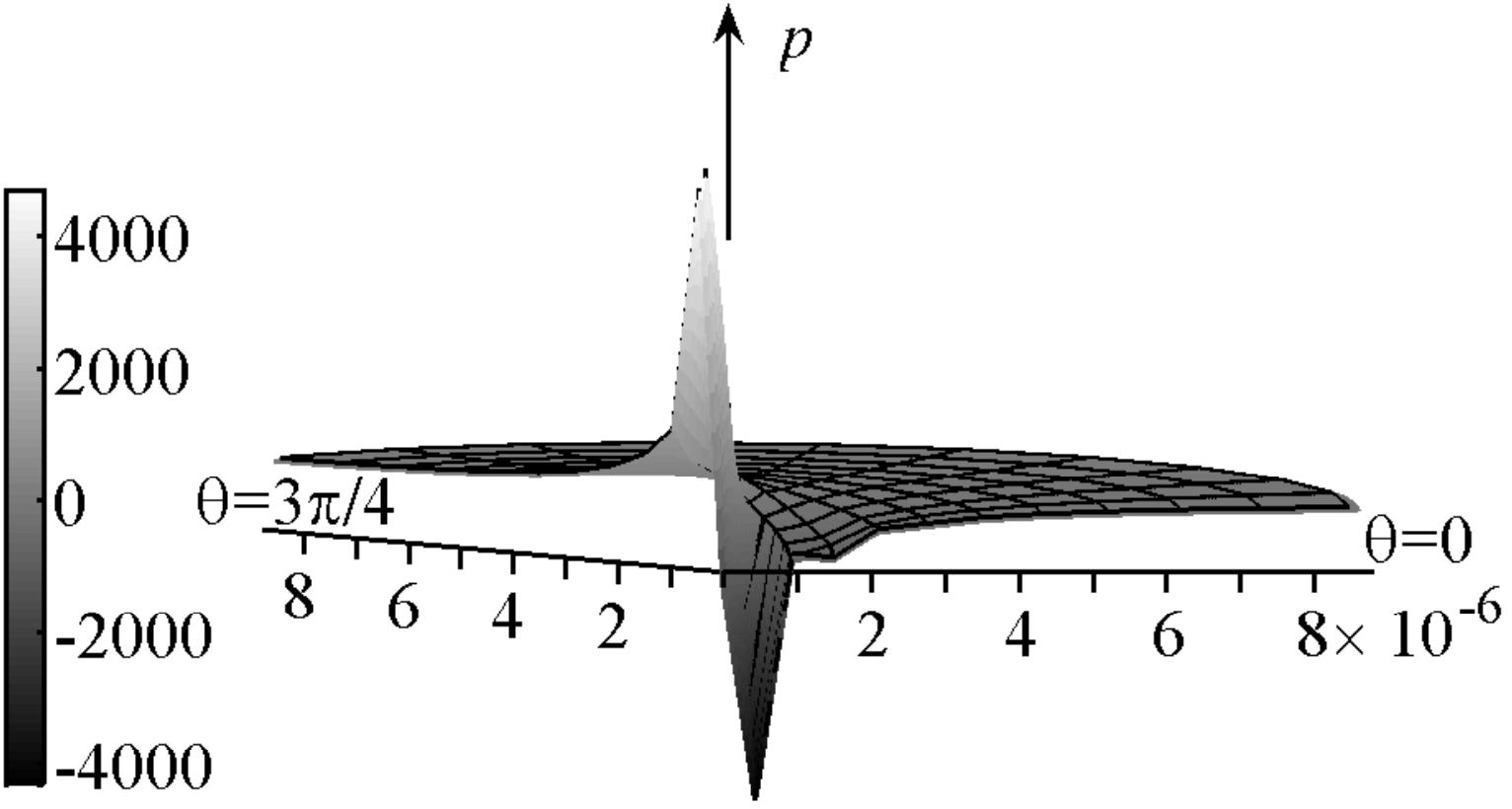}
\caption{Pressure distributions in the vicinity of the corner for $\alpha=3\pi/4$. Left: pressure
contours as the corner is approached with the underlying finite element mesh visible. Right:
surface plot of pressure.}\label{F:135eigpressure}
\end{figure}

\begin{figure}
\includegraphics*[viewport=-20 270 650 620,angle=0,scale=0.35]{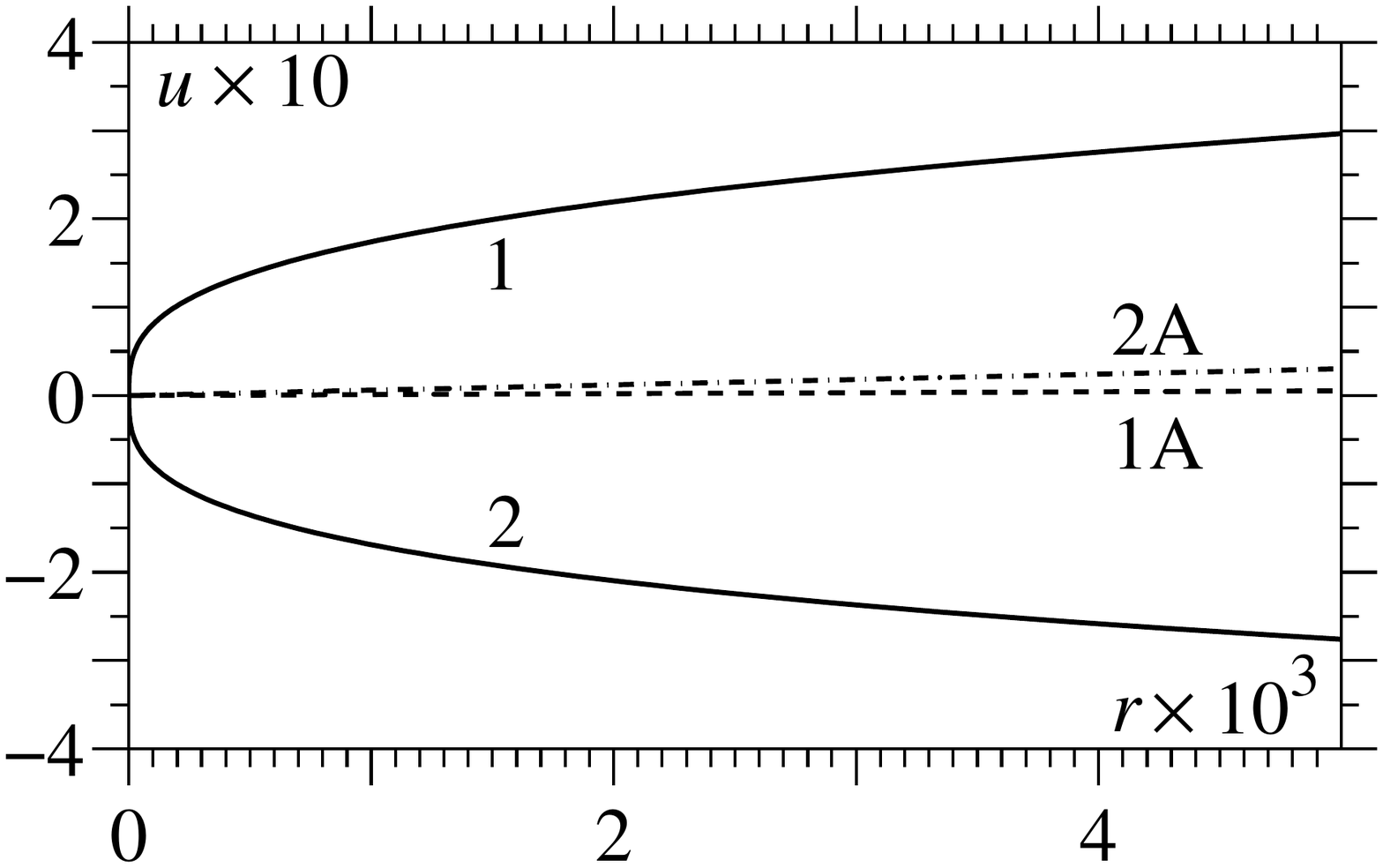}
\includegraphics*[viewport=-20 0 700 500,angle=0,scale=0.35]{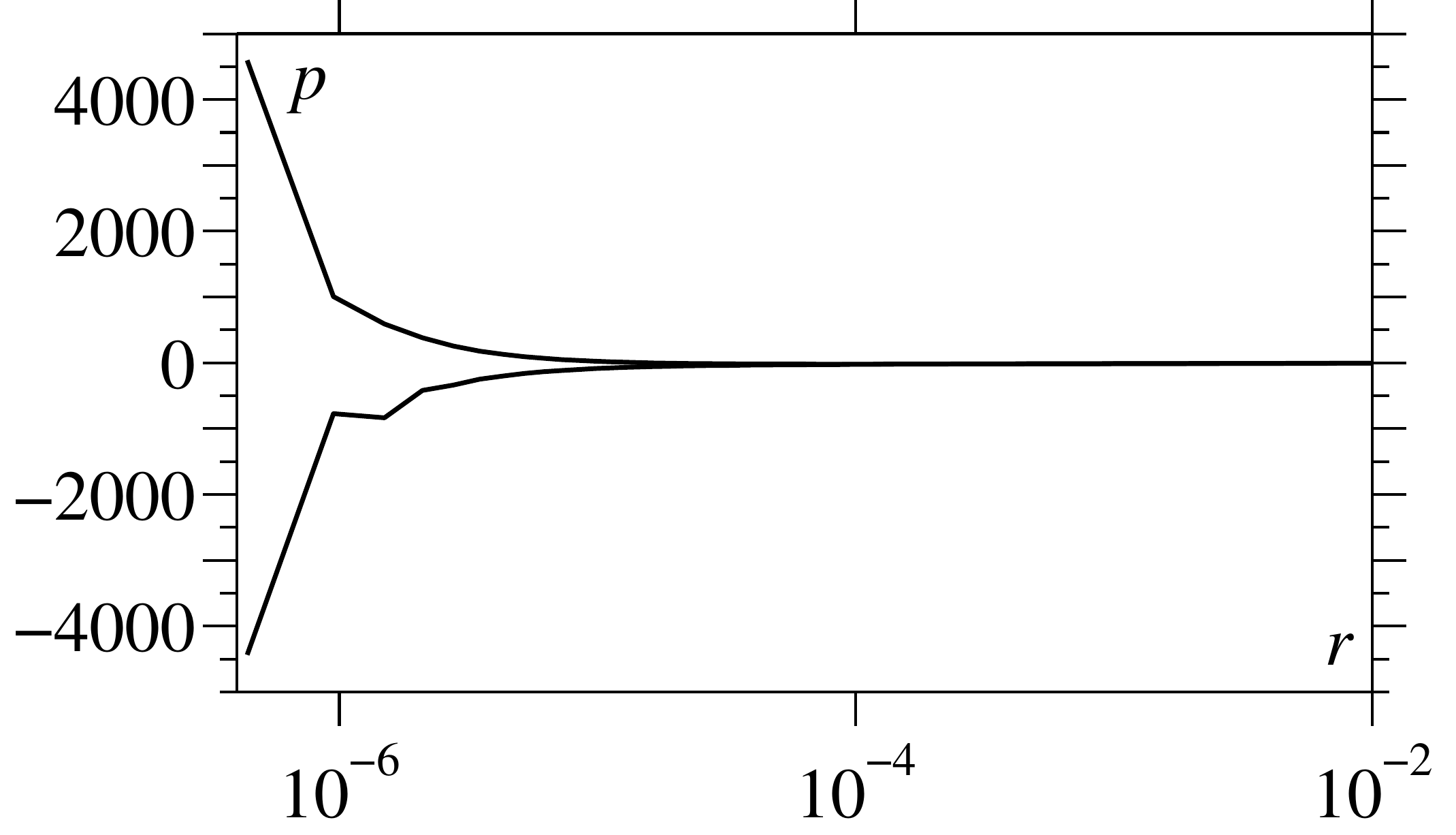}
\caption{Left: comparison of the computed radial velocity $u$ along the liquid-solid interface $1$
and liquid-gas interface $2$ with the corresponding asymptotic results $1A$ and $2A$, respectively.
Right: computed pressure along the interfaces.}\label{F:135eigasymptotics}
\end{figure}

Fig.~\ref{F:135eigstreamlines} shows the picture of streamlines near the corner. The flow is faster
than that obtained for acute angles but, at first sight, there is no indication of any particular
abnormality, at least on a qualitative level. However, when we examine the plots of the pressure
distribution near the corner shown in Fig.~\ref{F:135eigpressure}, it becomes clear immediately
that there are severe numerical issues.  The smooth $\theta$-independent pressure obtained for
acute angles is now replaced by two huge spikes of differing signs at the nodes adjacent to the
corner. The distributions of velocity and pressure along the boundaries near the corner shown in
Fig.~\ref{F:135eigasymptotics} confirm that we have a serious problem. The numerical method implies
single-valuedness for both velocity components and the pressure at the corner, and computations
confirm single-valuedness of velocity\footnote{Its single-valuedness is ensured by the slip
boundary condition \cite[see][]{dussan74,dussan79}, whereas, numerically, once we assume all
functions to be single-valued, the impermeability conditions make the corner a stagnation point.}
(Fig.~\ref{F:135eigasymptotics}). However, as one can see in Fig.~\ref{F:135eigasymptotics}, the
velocity differs drastically from what is predicted by the asymptotics, most noticeably by not
behaving linearly with radius. Furthermore, the asymptotics predicts that near the corner the flow
along the free surface is in the upstream direction with the radial component of velocity along the
free surface being positive. This is in stark contrast to the streamline pattern observed in
Fig.~\ref{F:135eigstreamlines}, which shows a regular downstream flow with the velocity on the free
surface directed towards the corner, as one may intuitively expect.

Although the problem with the computed velocity distribution is serious enough, the situation with
the pressure is even more severe. As one can see from Fig.~\ref{F:135eigpressure} and
Fig.~\ref{F:135eigasymptotics}, the numerical scheme is attempting to approximate a solution which
is both singular and multivalued at the corner. As one approaches the corner along different radii,
the pressure behaves differently  and tends to plus infinity along the radii closer to the free
surface and to minus infinity along those closer to the solid boundary. Since the numerical scheme
imposes (artificial) single-valuedness at the corner by stating that pressure has one value at the
corner node, the code tries to reconcile the calculated pressure with this requirement over the
first row of elements, thus creating the cliffs shown in Fig.~\ref{F:135eigpressure}. This makes
the numerical results mesh-dependent: the smaller one makes the size of the first row of elements,
the nearer the multivalued solution approaches the corner, and the higher the cliffs of pressure
become. Obviously, the mesh-dependent numerical results cannot be regarded as a uniformly valid
approximation of the actual solution to (\ref{contin})--(\ref{vect_far_field}) and hence have to be
rejected.

Furthermore, as shown by the semi-logarithmic plot in Fig.~\ref{F:135eigasymptotics}, the
singularity of pressure at the corner is stronger than logarithmic (by plotting the pressure in the
log-log frame one can show that it is actually algebraic with the exponent dependent on $\alpha$).
Besides being at odds with the asymptotics, such behaviour of pressure poses a fundamental problem
for computations of free-surface flows with moving contact lines for which the corner flow
considered here is but an element: a nonintegrably singular pressure distribution along the free
surface makes corrections to the wedge shape divergent and hence does not allow one to compute the
free surface shape at all \cite[see][]{shik07}.

It should be emphasized that the difficulties we are describing are above the level of a particular
numerical implementation of a particular algorithm. It is not only that the code we used has been
thoroughly validated (see the previous section); additionally, the commercially available code
COMSOL Multiphysics has also been applied to the wedge flow problem for both acute and obtuse
angles, and in all cases identical results have been recorded.

The encountered problem turns out to be resilient to standard approaches which have been described
in the literature as successful remedies to spurious numerical effects. The first approach to be
tried is to incorporate the asymptotic results described earlier in the numerical scheme to `come
out' of the corner, thus bypassing any spurious effects that might have been caused by the
geometry. The asymptotics can be matched with the numerical solution outside a given radius in a
variety of ways. This approach has proved successful in similar situations in which corner
singularities exist \cite[see][]{shi04}. However, it has been found that for our problem such an
alteration of the scheme merely shifts the pressure cliff to the arc at the radius where the
asymptotics has been applied. This is of no help whatsoever, since the results have all the
deficiencies listed above.

Another standard approach is to impose penalties of various forms, similar to those used with
equal-order interpolation to circumvent the LBB condition \cite[see][]{cairncross00a}, but such a
method simply fails to drive the code into a mesh-independent solution. The extent to which the
penalties flatten the pressure cliffs is exactly the extent to which the numerical solution (to the
`penalized problem') departs from that which would satisfy the original problem discretized using
the standard FEM. In other words, instead of driving the code to the `right' solution, the penalty
becomes part of the differential equations the code is solving, i.e.\ the method effectively
replaces the original problem with a different one.

The spurious pressure behaviour which we have shown throughout the paper has previously been
observed and designated the subject of future research in a paper on the finite element simulation
of curtain coating \cite[see][]{wilson01}.  As is pointed out in this paper, a reason that the
problem may have not been treated in other investigations is a lack of computational resolution.  A
rough indication of the region which must be resolved by any numerical scheme is the slip length,
given by $1/\bar{\beta}$. Tests show that at least $100$ nodes should be used along each radial-ray
within this region. With the graded mesh we have used and for modern computers this requirement
causes no problem. For a mesh with a single element size, achieving this is more difficult, as the
numerical cost of having every element of size $0.01/\bar{\beta}$ could render the computational
task untractable. It may be that this is the case in a number of older publications in which
computational power proved a major problem \cite[see][]{kistler83,christodoulou89}. In our
calculations presented here, the smallest element has the size of $4\times10^{-7}$, and we have
$206$ nodes inside the slip length along each radial-ray.


\section{\label{ori}Origin of the pressure multivaluedness}

At this stage, the question is whether the difficulty, albeit resilient to conventional alterations
of the algorithm, is a numerical artifact, or whether at $\alpha>\pi/2$ the computed features are a
genuine property of the mathematical problem. The robustness of the velocity distribution and the
fact that, contrary to the asymptotics, its radial dependence is strongly nonlinear, with a
singular radial derivative at the corner, suggest that here we might be dealing with an
eigensolution, i.e.\ a solution satisfying homogeneous boundary conditions, $v=\partial
u/\partial\theta=0$ $(\theta=0,\alpha)$, that superimposes on the one described by the asymptotics.
Then, if the eigensolution for the velocity has indeed singular radial derivatives at the corner,
the computed pressure behaviour could result from numerical errors in calculating the velocity
field corresponding to this eigensolution.

An eigensolution to our problem has been known for a long time \cite[see][P.14]{moffatt64}. In
terms of the stream function, it has the same separable form as the asymptotic solution described
in Section~\ref{ana}:
\begin{equation}
\label{eigensolution}
\psi_e=Ar^\lambda\sin(\lambda\theta),\qquad\lambda=\pi/\alpha,
\end{equation}
where $A$ is an arbitrary constant.

This solution is promising for two reasons. Firstly, it is only for $\alpha>\pi/2$ that we have
$\lambda<2$ and hence this solution can dominate the one described by the local asymptotics
(\ref{stream_2}) in the near field. Secondly, it exhibits the numerically computed non-linear
radial dependence of velocity. Note that, although the velocity of the eigensolution tends to zero
at the corner, its radial dependence scales like $r^{\lambda-1}$, and hence for $\pi/2<\alpha<\pi$
the derivatives of velocity in the radial direction are singular at the corner.

Crucially, the eigensolution (\ref{eigensolution}) produces a {\it globally constant\/} pressure.
This simplicity allows an ideal opportunity to check whether the nonintegrability and
multivaluedness of pressure computed earlier are indeed numerical artifacts.   In order to do this,
we consider the flow in a corner region with the Navier slip condition replaced with zero
tangential stress, i.e.\ with $\partial u/\partial\theta=0$ and $v=0$ on both boundaries $\theta=0$
and $\theta=\alpha$. The flow can then be generated only by the boundary conditions in the far
field which we set using the eigensolution (\ref{eigensolution}), where, for simplicity, we use
$A=1/\lambda$:
\begin{equation}\label{eig_far_field}
u=r^{\lambda-1}\cos(\lambda\theta),\quad
v=-r^{\lambda-1}\sin(\lambda\theta) \qquad\qquad (r=R,\
0<\theta<\alpha).
\end{equation}
Then the eigensolution is the exact global solution to our test problem. As for the pressure
distribution, once the pressure level has been set, say, to zero, then one will have $p\equiv0$ in
the whole domain.

\begin{figure}
\includegraphics*[viewport=0 0 450 250,angle=0,scale=0.5]{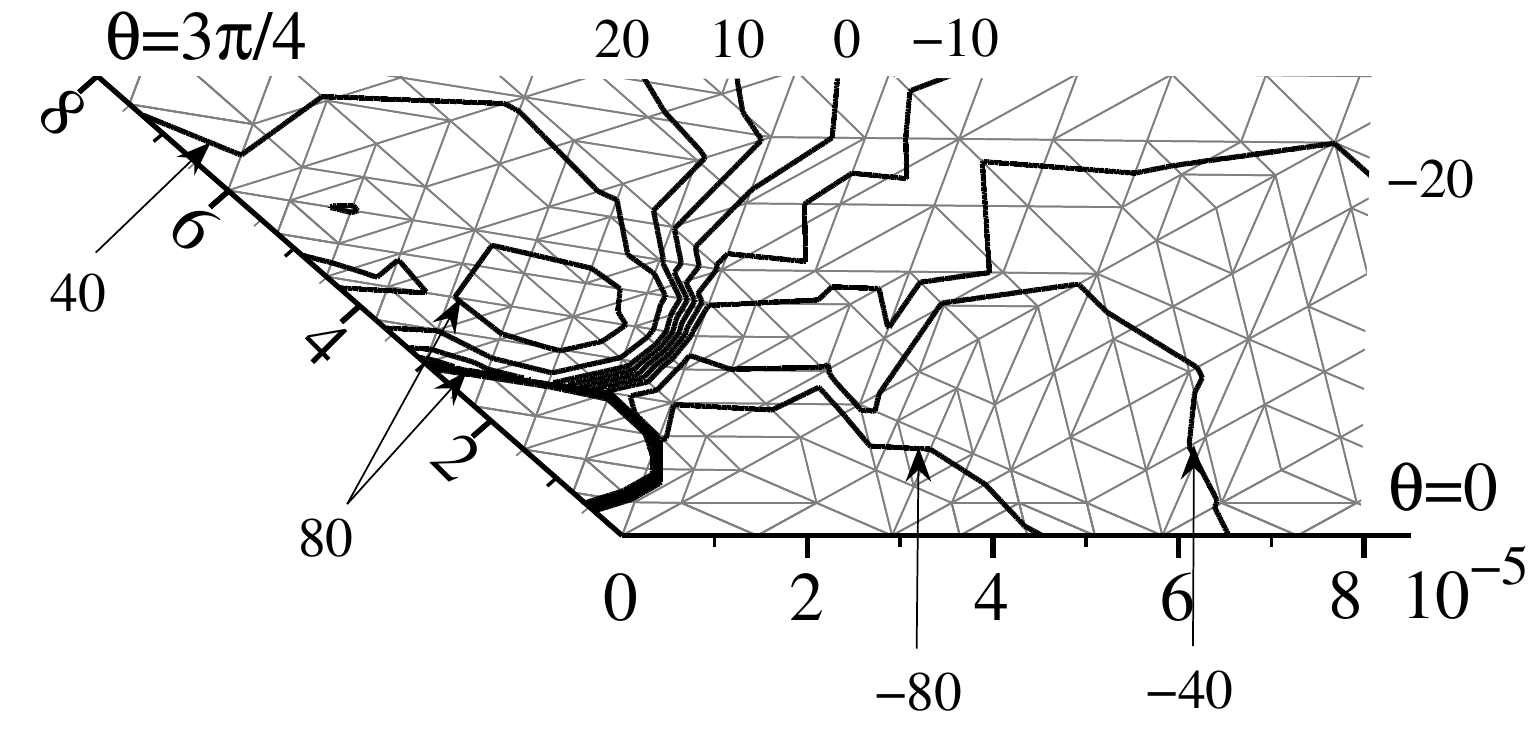}
\includegraphics*[viewport=0 -10 500 250,angle=0,scale=0.5]{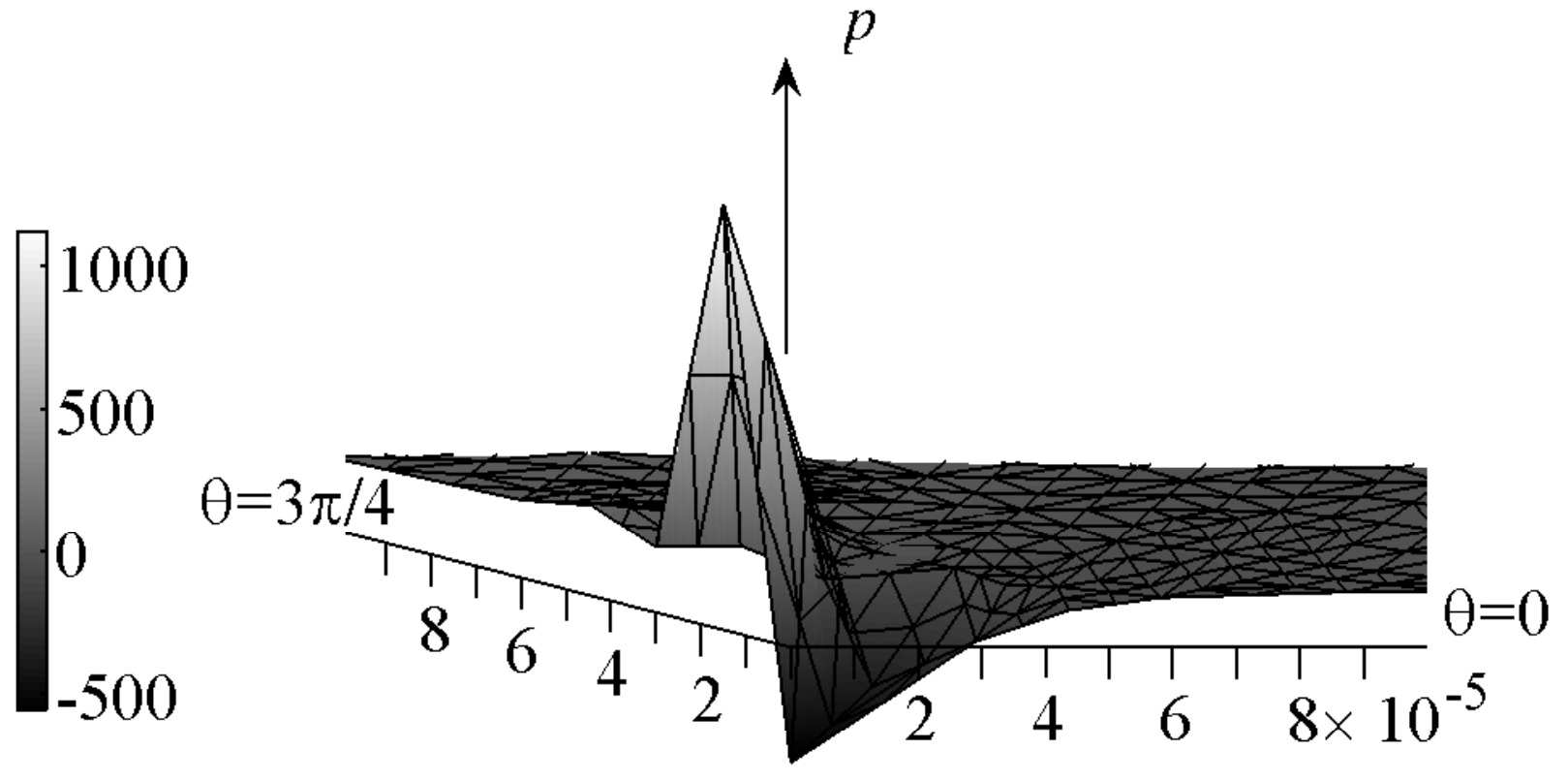}
\caption{Pressure distributions in the vicinity of the corner for $\alpha=3\pi/4$ as calculated by
COMSOL Multiphysics. The underlying finite element mesh is visible in both plots. Left: pressure
contours as the corner is approached. Right: surface plot of pressure.}\label{F:135comsolpressure}
\end{figure}

Once this problem is computed numerically, the situation becomes clear.
Fig.~\ref{F:135comsolpressure} shows the isobars and a 3-dimensional plot obtained using COMSOL
Multiphysics, again with the P2P1 element. The isobars and the 3-dimensional plot resemble those
which we have already seen for the Navier condition in the previous section, with the same cliffs
in the pressure profile at the nodes nearest to the corner.  This is a remarkable result given that
we know the global solution is in fact $p\equiv0$! The results obtained using our numerical
platform and those of COMSOL are in perfect quantitative agreement in the far field and are similar
in the near field; there is little point in comparing specific values in the near field as the
results are seen to be mesh dependent for both codes. An interested reader can easily reproduce the
results using COMSOL, where the condition of impermeability and zero tangential stress is selected
as a boundary condition by choosing boundary `wall' and then choosing the type `slip'.

The underlying mesh, visible for both plots in Fig.~\ref{F:135comsolpressure}, has been generated
by COMSOL using adaptive grid refinement techniques and it is unstructured. The same spurious
features of the numerical solution have been encountered for both the unstructured mesh of COMSOL
and the structured mesh of our numerical code, thus suggesting that any alternative mesh design
would not allow a route out of the problem. Additionally, COMSOL offers the choice of ten different
elements, and, as we have tested, the choice of any of these elements does not affect the overall
picture.

Therefore, we may now conclude that the pressure multivaluedness is
spurious and has a numerical origin.  In the next section, we will
show how the information we have about the eigensolution as the
underlying cause of the problem can be used to resolve it.

\section{\label{rem}Removal of pressure multivaluedness}

First, we will describe the method of removing the pressure multivaluedness from the numerical
solution of the problem formulated in Section 2, i.e.\ for the Stokes flow in a wedge region, and
then show how the method can be `localized' to apply to general 2-dimensional Navier-Stokes flows,
both steady and unsteady, where an angle formed with the solid surface by \emph{a priori} unknown
free boundaries is but one element.

The key idea of the method is very simple. In the situation where, as in our case, the
eigensolution is {\it the\/} cause of the pressure multivaluedness, we can subtract this solution
from the problem and use the degree of freedom it offers, i.e.\ arbitrariness of $A$ in
(\ref{eigensolution}), to ensure single-valuedness of pressure at the corner. Once the
supplementary to the eigensolution velocity and the pressure fields are computed, we can put the
eigensolution, i.e.\ the velocity field described by (\ref{eigensolution}) and uniformly zero
pressure, back in. The resulting solution will have the analytically known eigencomponent of the
velocity field superimposed on the computed `supplementary' flow and the computed single-valued
pressure. The resultant combined solution will satisfy the original equations and boundary
conditions.

\subsection{\label{6.1}Simplest case: Stokes flow in a corner region}

For the problem formulated in Section~\ref{pf}, consider the
velocity and pressure as sums of the eigensolution
\begin{subeqnarray}
\label{eigen2}\gdef\thesubequation{\theequation \textit{a,b,c}} u_e=A\lambda
r^{\lambda-1}\cos(\lambda\theta), \quad v_e=-A\lambda r^{\lambda-1}\sin(\lambda\theta),\quad p_e=0,
\qquad (\lambda=\pi/\alpha),
\end{subeqnarray}
and the components to be computed (hereafter these are marked with a tilde):
\begin{equation}\label{vel_super}
(u,v,p) = (u_e,v_e,p_e) + (\tilde{u},\tilde{v},\tilde{p}).
\end{equation}
The constant $A$ in (\ref{eigen2}\textit{a,b}) is yet to be specified.

Since the eigensolution satisfies the Stokes equations (\ref{contin})--(\ref{motion_prim}) and the
free-surface boundary conditions (\ref{vect_fs}) exactly, one has that $\tilde{u}$, $\tilde{v}$ and
$\tilde{p}$ have to satisfy the unaltered Stokes equations (\ref{contin})--(\ref{motion_prim}) and
boundary conditions (\ref{vect_fs}), whereas the boundary conditions on the solid surface and in
the far field for these variables will take the form:
\begin{subeqnarray}\label{vect_ss_new}
\gdef\thesubequation{\theequation \textit{a,b}} \frac{\partial\tilde{u}}{\partial\theta}
=r\bar{\beta}\left(A\lambda r^{\lambda-1} + \tilde{u}-1\right), \qquad \tilde{v}=0 \qquad (0<r<R,\
\theta=0),
\end{subeqnarray}
\begin{subeqnarray}\label{vect_ff_new}
\frac{\partial\tilde{u}}{\partial r}=-A\lambda(\lambda-1) r^{\lambda-2}\cos(\lambda\theta),\qquad
\frac{\partial\tilde{v}}{\partial r}=A\lambda(\lambda-1) r^{\lambda-2}\sin(\lambda\theta), \qquad
(r=R,\ 0<\theta<\alpha).
\end{subeqnarray}
To complete the problem formulation for $\tilde{u}$, $\tilde{v}$ and $\tilde{p}$, we must add an
equation to account for the additional unknown constant $A$.  To do so, we impose a condition that
the pressure is single valued at the corner:
\begin{equation}\label{pextra}
\lim_{r\rightarrow 0} \frac{\partial\tilde{p}}{\partial\theta}=0.
\end{equation}
Qualitatively, this condition can be explained as follows. The method is based on taking the
eigensolution out of the total and hence ensuring that $(\tilde{u},\tilde{v})$ do not have the
singularity of the radial derivatives at the corner, and the numerical error in their computation
will not give rise to the errors in computations of the pressure resulting in its multivaluedness.
By imposing (\ref{pextra}), we are effectively ensuring that the eigensolution is taken out {\it
fully\/} from the viewpoint of what this subtraction is aimed at achieving. In other words, out of
a one-parametric family of eigensolutions, parameterised by the constant $A$, condition
(\ref{pextra}) selects the one that underpins the flow we are considering.

A simple way for us to impose (\ref{pextra}) numerically is to demand that the pressures at the
nearest to the corner nodes on the free surface and on the solid boundary are equal, as in our mesh
they are equidistant from the corner.

\begin{figure}
\centering
\includegraphics*[viewport=-100 0 800 300,angle=0,scale=0.7]{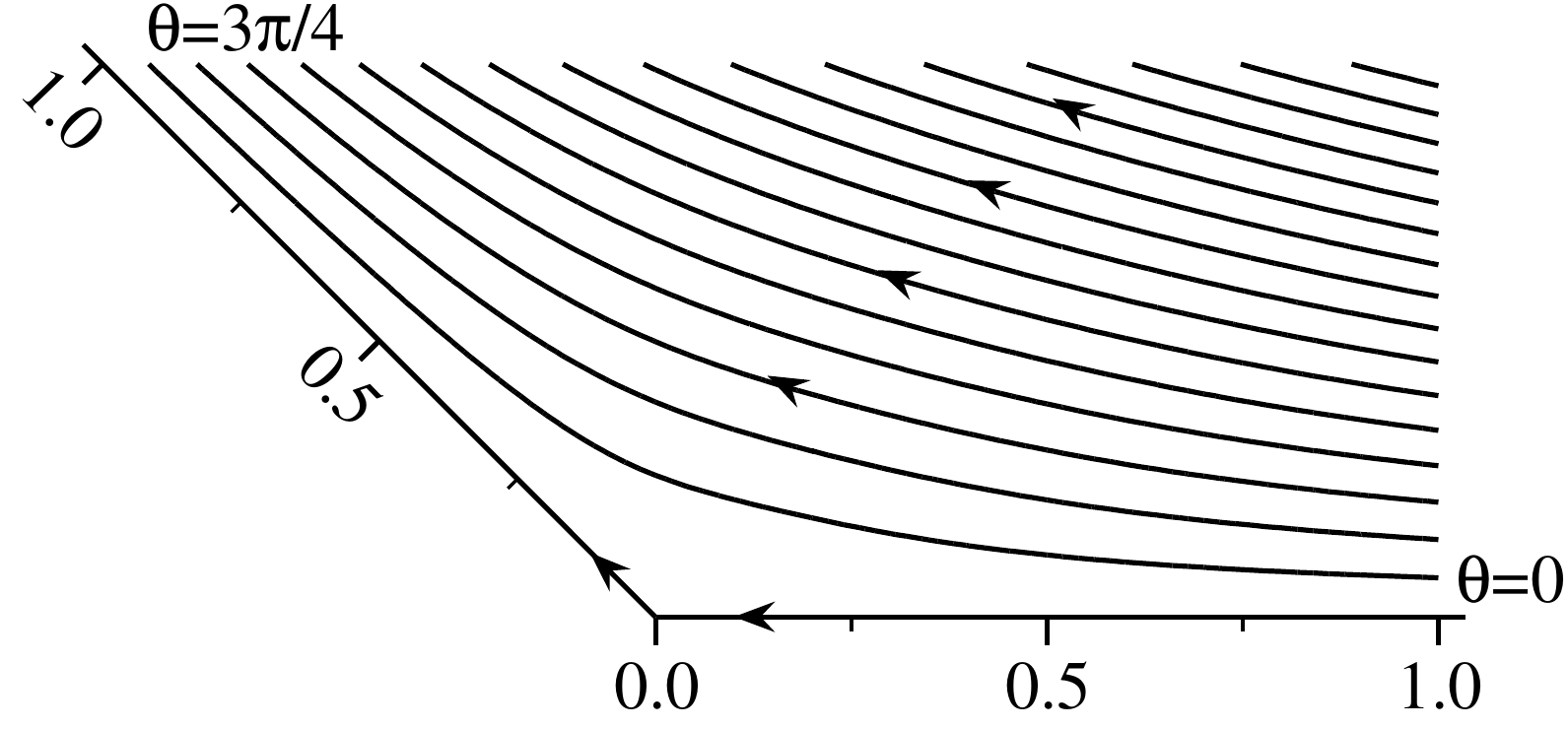}
\includegraphics*[viewport=-100 0 800
300,angle=0,scale=0.7]{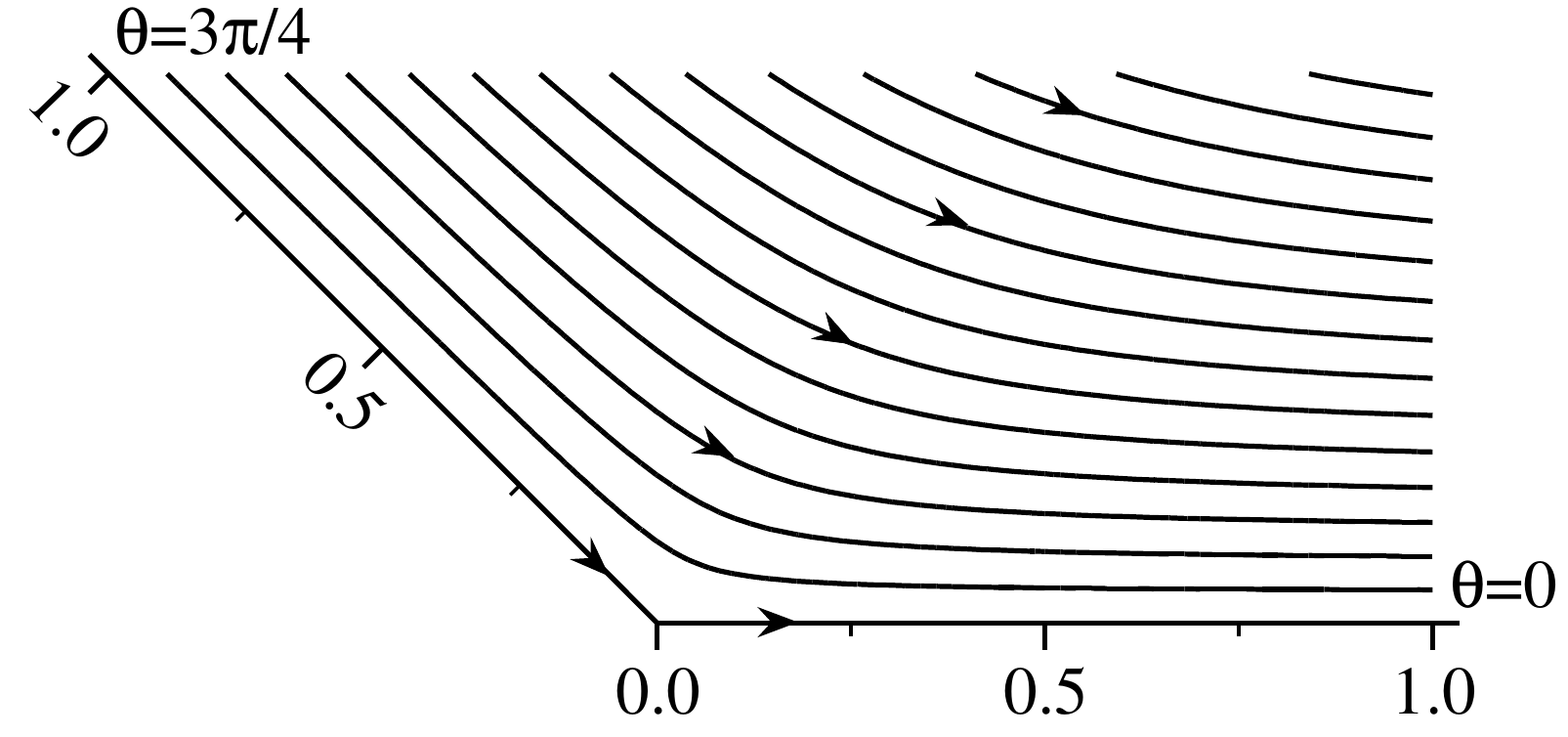} \caption{Top: the computed
streamlines in increments of $\psi=-0.04$. Bottom: the streamlines with the eigensolution
superimposed to produce the full solution with increments of $\psi=0.04$.  In both plots
$\alpha=3\pi/4$  with the boundaries at $\theta=0$ and at $\theta=\alpha$ corresponding to
$\psi=0$. The constant $A = 1.3026$.} \label{F:135fixedstreamlines}
\end{figure}

Equations (\ref{contin})--(\ref{motion_prim}), (\ref{vect_fs}),
(\ref{vect_ss_new})--(\ref{pextra}) have been solved using our
numerical platform with the same solution procedure as before. The
streamlines of the supplementary flow obtained from the
computation are shown in Fig.~\ref{F:135fixedstreamlines}
alongside the streamlines to our original problem formulated in
Section 2, obtained using (\ref{vel_super}). Although the
underlying asymptotic solution predicts that there must be flow
{\it reversal\/} near the corner, which is replicated in our
numerical solution for $(\tilde{u},\tilde{v})$, this feature is
blown away by the strength of the eigensolution when it is
superimposed on top.

\begin{figure}
\includegraphics*[viewport=100 0 600 250,angle=0,scale=0.5]{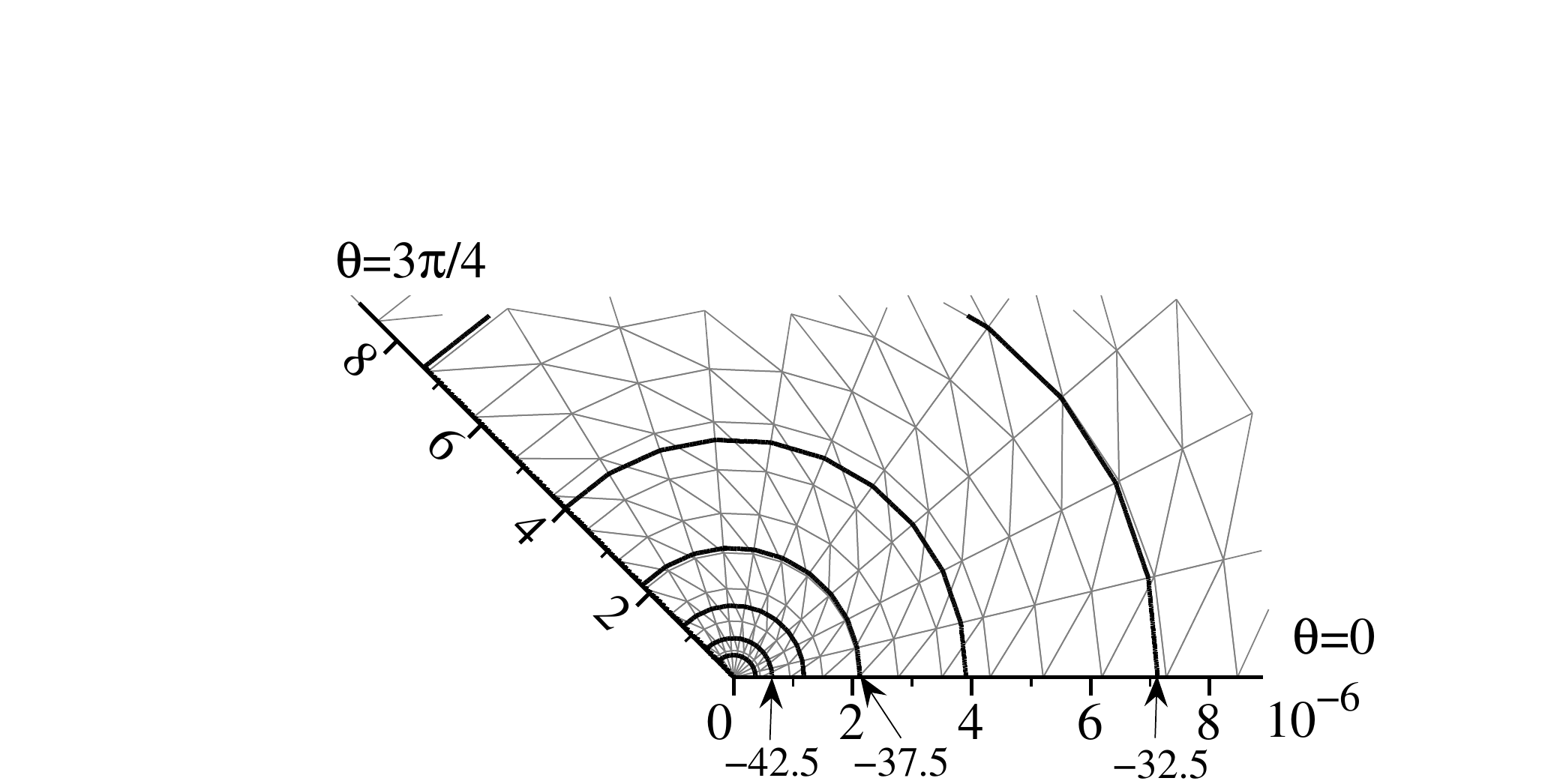}
\includegraphics*[viewport=0 0 550 300,angle=0,scale=0.5]{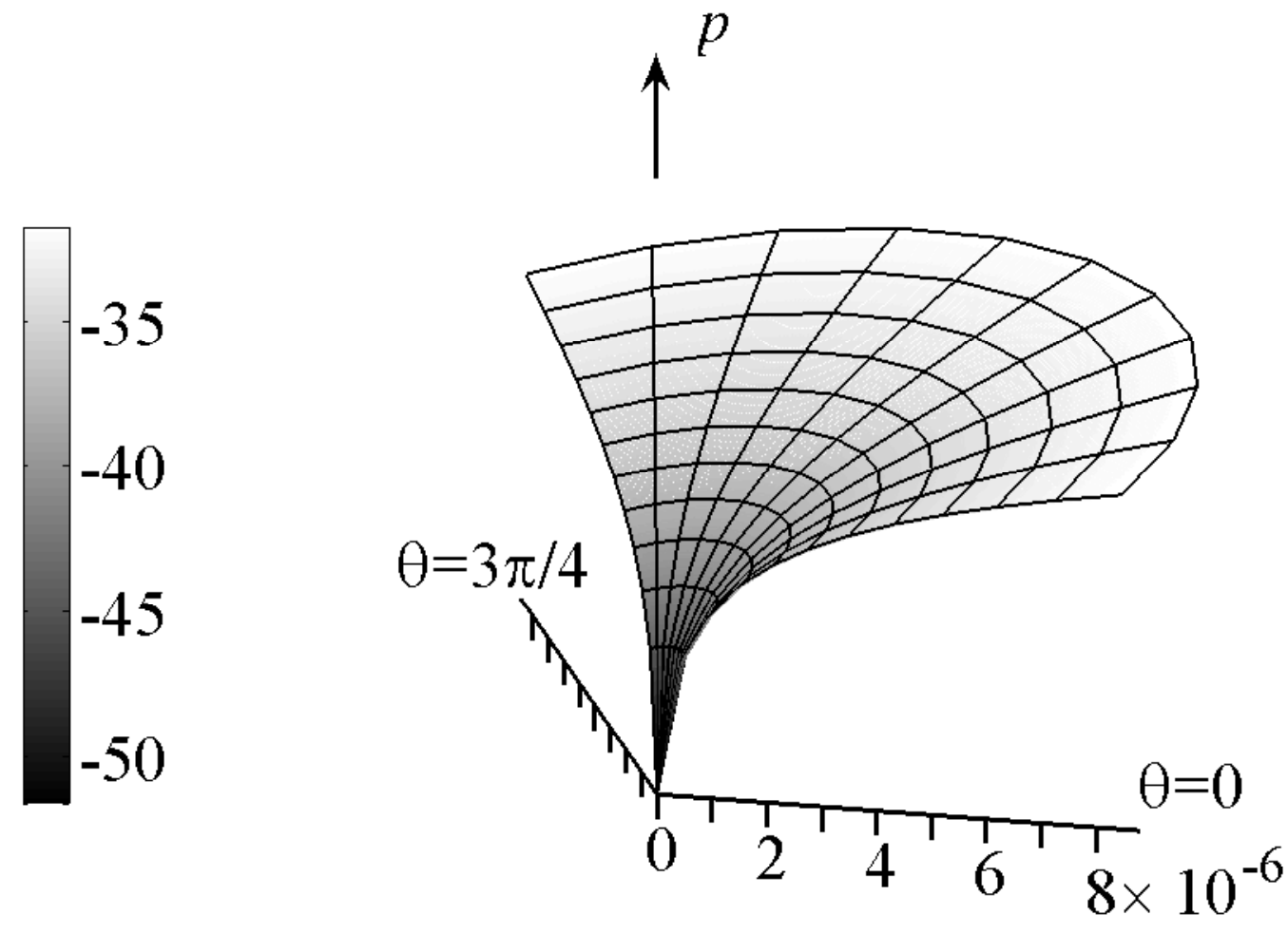}
\caption{Pressure distributions in the vicinity of the corner for $\alpha=3\pi/4$ using our
numerical platform with the eigensolution removed and $A=1.3026$. Left: pressure contours in steps
of size 2.5 as the corner is approached, with the underlying finite element mesh visible. Right:
surface plot of pressure.}\label{F:135fixedpressure}
\end{figure}

The pressure plots in Fig.~\ref{F:135fixedpressure} show a qualitative transformation from those
observed in previous sections: the pressure is now single valued, smooth and exhibits no
mesh-dependence. Both the isobars and the 3D `surface plot' are now similar to those obtained
earlier for acute angles (Fig.~\ref{F:45pressure}).

\begin{figure}
\includegraphics*[viewport=-20 270 700 620,angle=0,scale=0.35]{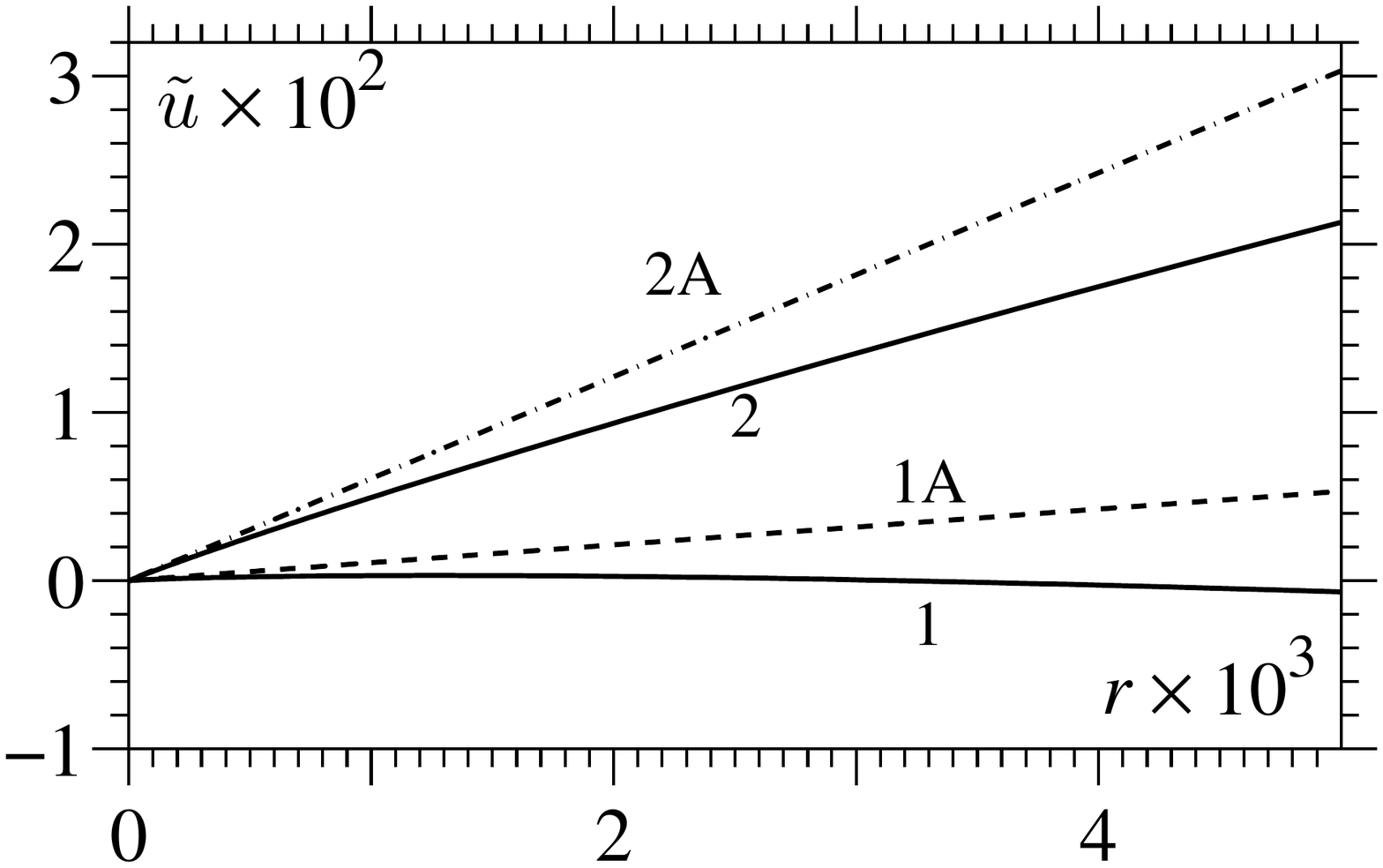}
\includegraphics*[viewport=-20 270 700 620,angle=0,scale=0.35]{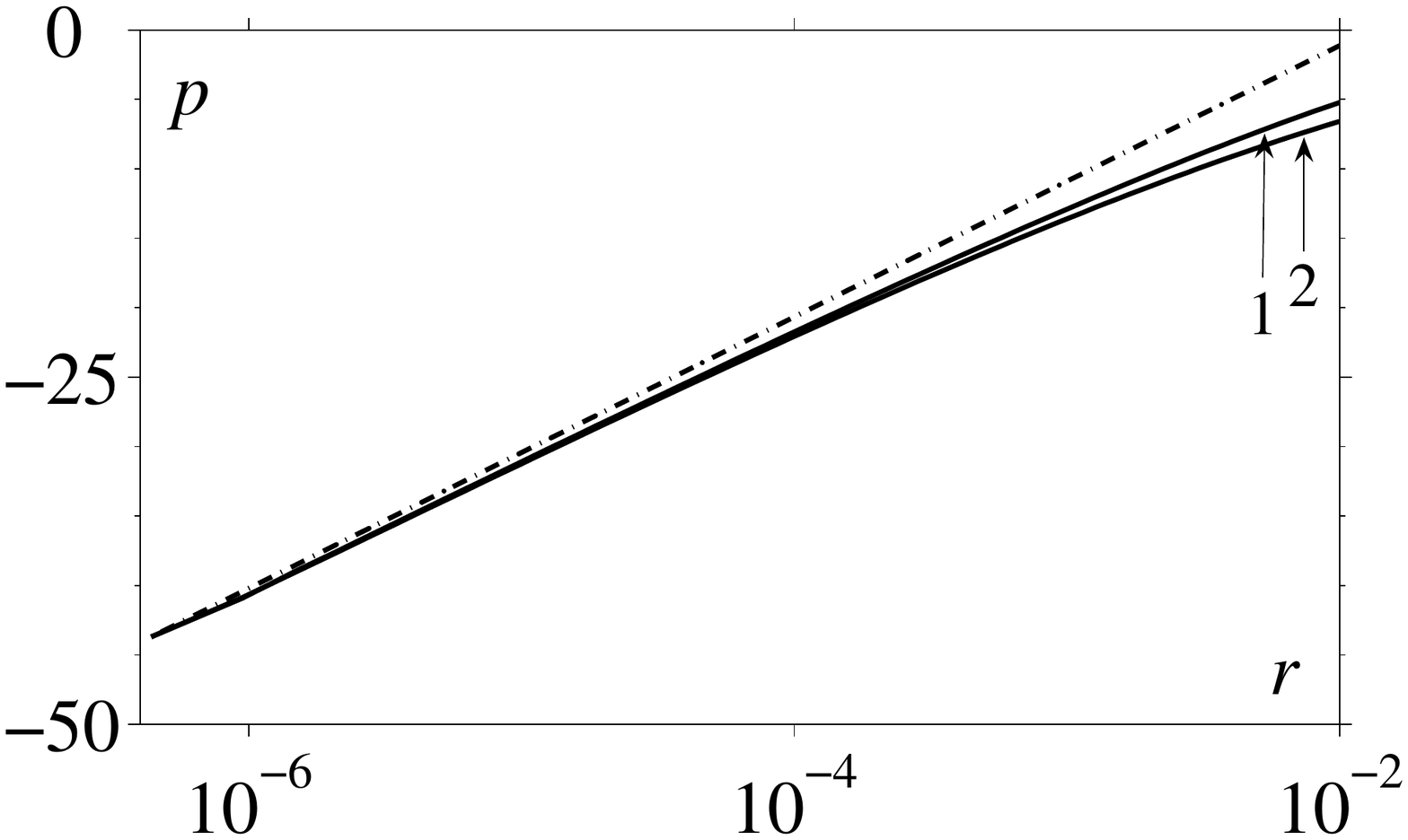}
\caption{Left: comparison of the computed radial velocity $\tilde{u}$ along the liquid-solid
interface $1$ and liquid-gas interface $2$ with the asymptotic results $1A$ and $2A$, respectively.
Right: comparison of the computed/actual pressure along the interfaces to the asymptotic result
(dashed line).}\label{F:135fixedasymptotics}
\end{figure}

The comparison with the asymptotics of Section 3 is shown in Fig.~\ref{F:135fixedasymptotics}. The
agreement of pressure is visibly excellent.  The fit between the asymptotics and the radial
velocity along the interfaces, $\tilde{u}$, is now confined to a smaller region than that
previously observed for acute angles. This is no surprise, given that the eigensolution now enters
the Navier conditions (\ref{vect_ss_new}\textit{a}), where the first term in brackets on the
right-hand side becomes small compared to the last term only very close to the corner.

Interestingly, the dominance of the eigensolution in the combined solution for obtuse angles
ensures that the velocity field near the corner is almost anti-symmetric about the centre line
$\theta=\alpha/2$, whereas for acute angles this is not the case.


Thus, for the present problem of flow in a corner, we have fully resolved the situation.  However,
when considering more complicated flows which involve a corner only as an element, the method of
removing the eigensolution from the problem formulation throughout the entire domain could become
unnecessarily complicated; the eigensolution is only important near one corner and yet it will make
artificial contributions to equations and boundary conditions in the whole domain. A more
reasonable approach would be to design a `local' variant of the method to remove the eigensolution
only near the corner, where its presence creates the unwanted numerical artifacts, leaving the rest
of the flow domain intact. This variant of the method is considered below.

\subsection{\label{6.2}General case: Navier-Stokes equations
in domains with curvilinear free boundaries}

We will begin by describing the pressure regularization method in its generic form and then
illustrate it by considering two test problems: (a) a steady propagation of a meniscus in a channel
with plane-parallel walls and (b) an unsteady spreading of a (2-dimensional) liquid drop over a
solid surface. In both cases, we will be describing only the details related to the implementation
of the method and leave out the standard elements of the problem formulation for these free-surface
flows.

Now, we have that the bulk flow is described by the full Navier-Stokes equations up to the corner
(which, in accordance with the literature, we will now call the `contact line') and the position of
the curved free surface is \emph{a priori} unknown.

In order to `localize' the pressure regularization method, we now introduce an artificial `internal
boundary' that will separate the `inner' region near the contact line where the regularization
procedure will be used from the rest of the flow domain (the `outer region'). Inside the inner
region, we remove the eigensolution prior to computing the supplementary components of the velocity
and pressure, and then superimpose it back, as we did in the previous subsection: this prevents the
spurious pressure behaviour described in Sections 4 and 5.  In the outer region, there is no need
to alter the standard numerical approach to solving the appropriate equations. Solutions in both
the inner and the outer region are computed simultaneously and have to be matched at the internal
boundary.

Thus, after decomposing the solution in the inner region into the
sum of the eigensolution and the supplementary part to be
computed,
\begin{equation}\label{local_vel_super}
(\bd{u},p) = (\bd{u}_{e},0) + (\tilde{\bd{u}},\tilde{p}),
\end{equation}
we need to:
\begin{itemize}
\item[(i)]
Take into account, as we did in \S\ref{6.1}, the contribution
of the eigensolution to the boundary conditions for
$\tilde{\bd{u}}$ on the solid surface.

\item[(ii)]
Take into account the contribution of the eigensolution to the
bulk equations for $\tilde{\bd{u}}$ and $\tilde{p}$ in the
inner region arising from the fact that the eigensolution
satisfies the Stokes but not the Navier-Stokes equations.

\item[(iii)]
Take into account the contribution of the eigensolution to the boundary conditions for
$\tilde{\bd{u}}$ and $\tilde{p}$ on the free surface that appears due to the fact that now the free
surface, now being genuinely free, is not necessarily planar.

\item[(iv)] Formulate the matching conditions at the internal
boundary that would link the $\tilde{\bd{u}}$ and $\tilde{p}$ with
the outer flow. These conditions are necessary to calculate
solutions in both the inner region and the outer region; these
calculations are carried out simultaneously.
\end{itemize}

After non-dimensionalizing the problem using characteristic length
$L$ and velocity $U$ scales from the global flow, on the flat solid
surface, which moves at non-dimensional speed $\bd{U}_w$, we have
the same Navier-slip and impermeability conditions as in
\S\ref{6.1}:
\begin{subeqnarray}\label{local_solid}\gdef\thesubequation{\theequation \textit{a,b}}
\bd{n}\cdot\tilde{\bd{P}}\cdot\left(\bd{I}-\bd{n}\bd{n}\right) = \bar\beta\left(\tilde{\bd{u}} +
\bd{u}_{e}-\bd{U}_w\right)\cdot\left(\bd{I}-\bd{n}\bd{n}\right),\qquad \left(\tilde{\bd{u}} +
\bd{u}_{e}-\bd{U}_w\right)\cdot\bd{n}=0,
\end{subeqnarray}
where $\tilde{\bd{P}} = -\tilde{p}\bd{I} + \left[\nabla\tilde{\bd{u}} +
\left(\nabla\tilde{\bd{u}}\right)^{T}\right]$ is the stress tensor of the supplementary solution;
$\bd{I}$ is the metric tensor, $\bd{n}$ is the internal normal to the interface, so that the tensor
$\left(\bd{I}-\bd{n}\bd{n}\right)$ extracts the tangential component of a vector.  In
(\ref{local_solid}\textit{a}), we have used that for the stress tensor of the eigensolution,
\begin{equation}\label{local_stress_super}
\bd{P}_e = \left[\nabla\bd{u}_e +
\left(\nabla\bd{u}_e\right)^{T}\right],
\end{equation}
one has $\bd{n}\cdot\bd{P}_{e}\cdot\left(\bd{I}-\bd{n}\bd{n}\right)=0$.

The bulk equations for $\tilde{\bd{u}}$ and $\tilde{p}$ to be
solved in the inner region take the form
\begin{subeqnarray}\label{local_bulk}  \gdef\thesubequation{\theequation \textit{a,b}}
\nabla\cdot\tilde{\bd{u}}=0,\qquad Re\left[\frac{\partial\left(\tilde{\bd{u}} +
\bd{u}_{e}\right)}{\partial t}+\left(\tilde{\bd{u}} +
\bd{u}_{e}\right)\cdot\nabla\left(\tilde{\bd{u}} + \bd{u}_{e}\right)\right] = \nabla\cdot
\tilde{\bd{P}}+St\,\hat{\bd{g}},
\end{subeqnarray}
where $St = \rho L^{2} g/(\mu U)$ is the Stokes number based on the acceleration due to gravity $g$
and $\hat{\bd{g}}$ is a unit vector in the direction of the gravitational force. In equations
(\ref{local_bulk}\textit{a,b}), we have used that the eigensolution satisfies the Stokes equations,
i.e.\ $\nabla\cdot\bd{u}_{e}=\nabla\cdot\bd{P}_{e} = 0$.

On the free surface, which is defined by $f(\bd{x},t)=0$, where
$\bd{x}$ is the position vector and the function $f$ is to be
found, we have the standard dynamic and kinematic conditions:
\begin{subeqnarray}\label{local_fs}\gdef\thesubequation{\theequation \textit{a,b}}
Ca\,\bd{n}\cdot\left(\tilde{\bd{P}} + \bd{P}_{e}\right)= \bd{n}\nabla\cdot\bd{n},\qquad
\frac{\partial f}{\partial t}+\left(\tilde{\bd{u}} + \bd{u}_{e}\right)\cdot\nabla f = 0,
\end{subeqnarray}
where $Ca=\mu U/\sigma$ is the capillary number based on the
constant surface tension coefficient $\sigma$ of the liquid-gas
interface.

In the numerical implementation, care must be taken in the evaluation of the term
$\bd{n}\cdot\bd{P}_{e}$ in (\ref{local_fs}\textit{a}) as, although it is integrable, it is singular
in the limit $r\rightarrow 0$.  The term is best evaluated by calculating the stress tensor
analytically, rather than using the finite element approximation, or whatever other discretization
has been chosen, for the derivatives of the eigensolution's velocity components in
(\ref{local_stress_super}).

An internal boundary separating the inner region from the outer flow should lie sufficiently far
away from the contact line for the eigensolution to be well within the inner region and at the same
time not too far for the regularization method to be localized, as opposed to applied to an
unnecessarily large region of the overall flow. Another consideration is how the position of the
internal boundary correlates with the computational mesh. In our numerical method this is most
easily achieved by using one of the arcs formed by the edges of the finite elements; the method is
equally applicable for an algorithm with an unstructured mesh, though the ease of defining the
internal boundary will be lost.

At the internal boundary, we enforce continuity on the velocity
and the stress:
\begin{subeqnarray}\label{local_trans}  \gdef\thesubequation{\theequation \textit{a,b}}
\bd{u}_{e} + \tilde{\bd{u}} = \bd{u}_{out},\qquad \bd{n}_i\cdot\left(\bd{P}_{e} +
\tilde{\bd{P}}\right) = \bd{n}_i\cdot\bd{P}_{out},
\end{subeqnarray}
where $\bd{n}_i$ is a normal to the internal boundary and the
subscript {\it out\/} marks the velocity and stress in the outer
region that also have to be computed.

The ability of the finite element method to incorporate stress
boundary conditions in a natural manner ensures that the conditions
of continuity on the actual solution (\ref{local_trans}) can be
satisfied without dropping any other equations.
\begin{figure}
\centering
\includegraphics*[viewport=-100 0 800 300,angle=0,scale=0.7]{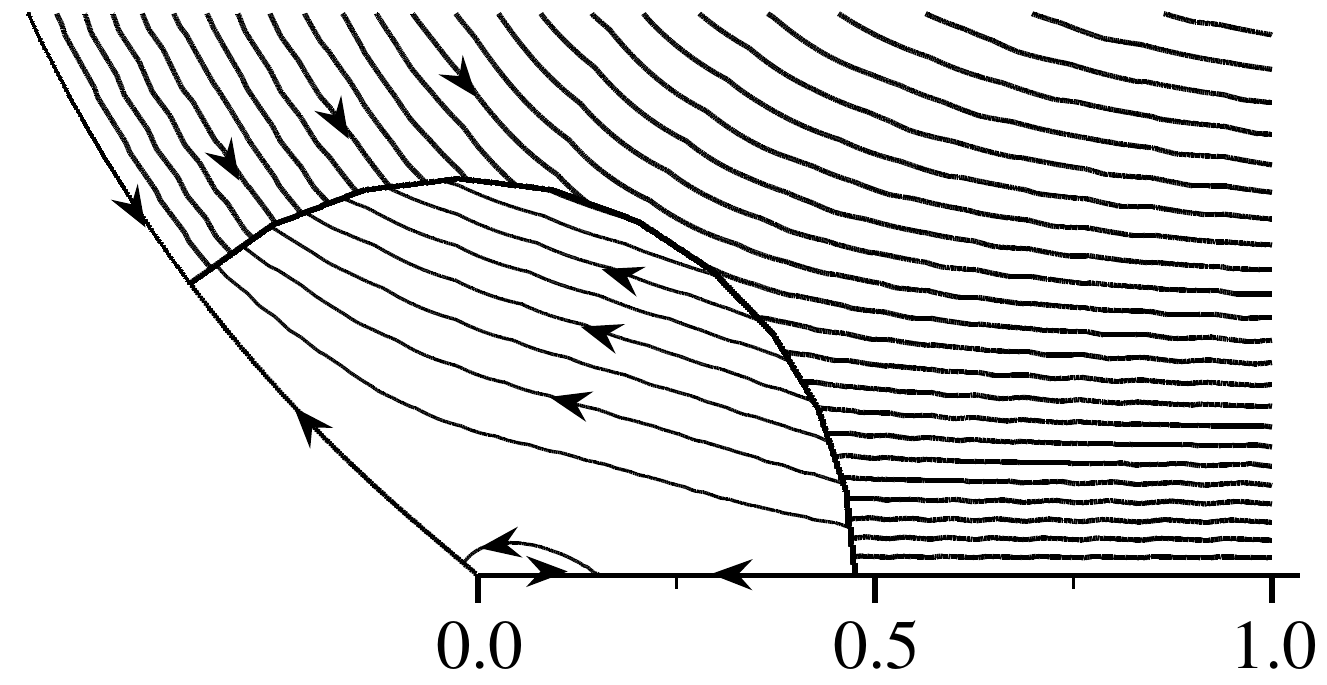}
\includegraphics*[viewport=-100 0 800 300,angle=0,scale=0.7]{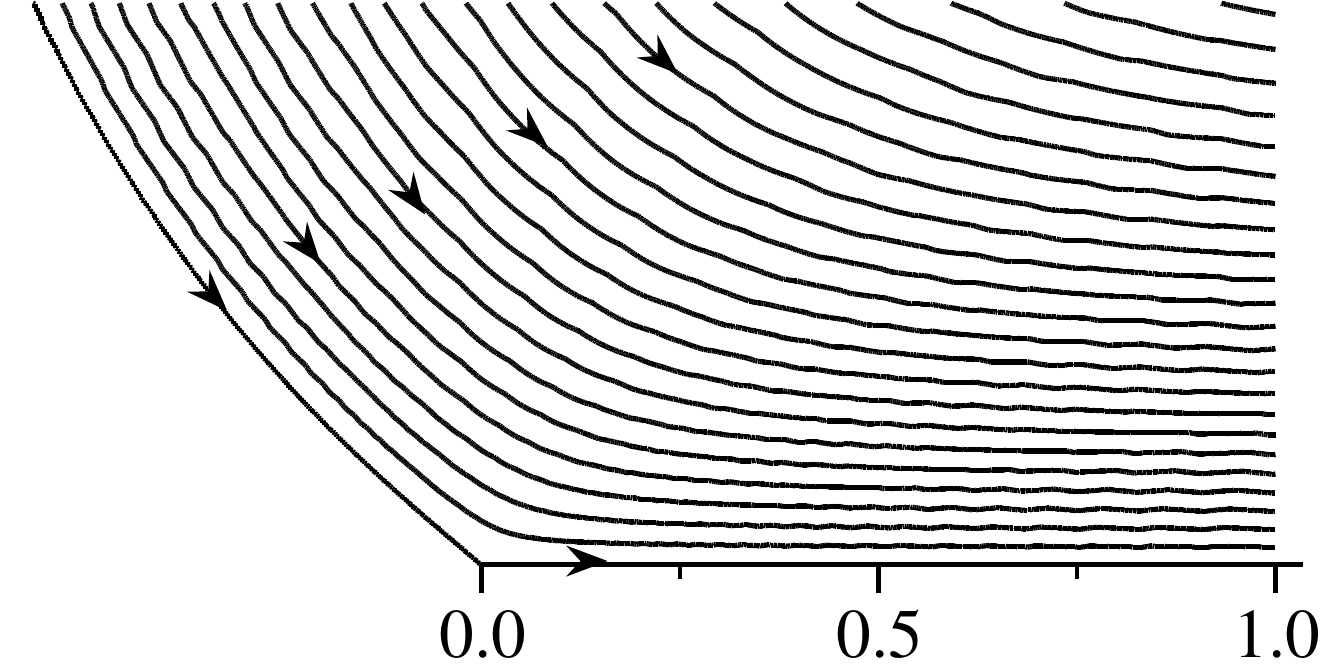}
\caption{Streamlines near the contact line of a propagating meniscus. The free surface and solid
surface both correspond to $\psi=0$. $A$ is computed to be $1.4551$. Top: streamlines of the
calculated velocity field with increments of $\psi=-0.02$ in the inner region and $\psi=0.02$ in
the outer region. Bottom: plot showing the actual streamlines obtained after superimposing the
eigensolution with the calculated solution in the inner region.} \label{F:135transstreamlines}
\end{figure}

In order to illustrate how the method works, we consider two model two-dimensional simulations
using a Cartesian frame $\bd{x}=(x,y)$ with a solid surface at $y=0$. In our simulations, we fix
the parameters to $Ca=0.1$, $Re=1$, $St=1$ and $\bar{\beta}=10$. As our interest here lies in the
numerical approximation of these flows, as opposed to a detailed comparison with experiment, we may
consider, for simplicity, the dynamic contact angle, which was previously labelled $\alpha$, to
have a fixed value of $\theta_{d} = 3\pi/4$.

The first problem is the steady motion of a meniscus that a fluid-gas interface forms between
plane-parallel walls.  In a frame moving with the contact line the problem is time-independent and
hence the time derivatives in (\ref{local_bulk}\textit{b}) and (\ref{local_fs}\textit{b}) are zero.
The velocity of the solid substrate in the moving reference frame is $\bd{U}_w=(1,0)$.

Fig.~\ref{F:135transstreamlines} shows the velocity fields as they have been computed and the
resulting (combined) velocity field. In the top picture, we can see the outer flow and a much
weaker supplementary flow in the inner region. When the eigensolution is superimposed back on top
of the supplementary velocity field in the inner region, the streamlines are seen to not feel the
presence of the internal boundary --- the matching conditions work perfectly leaving no `scar' on
the flow. As one can observe, peculiarities of the underlying (supplementary) flow in the inner
region, such as flow reversal, are of little consequence as their effect is negligible compared to
that of the eigensolution.

\begin{figure}
\centering
\includegraphics*[viewport=100 30 600 300,angle=0,scale=0.7]{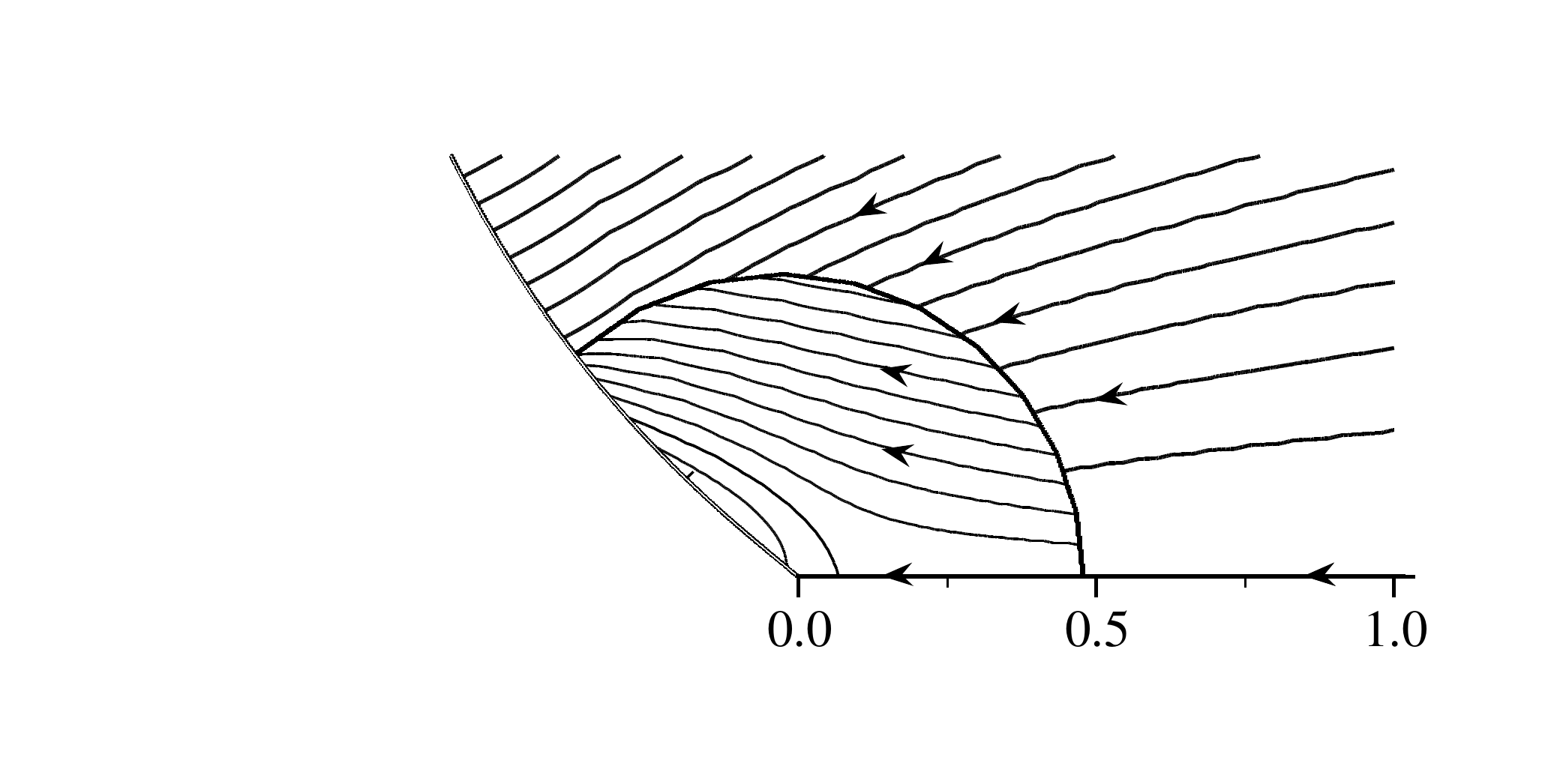}
\includegraphics*[viewport=100 30 600 300,angle=0,scale=0.7]{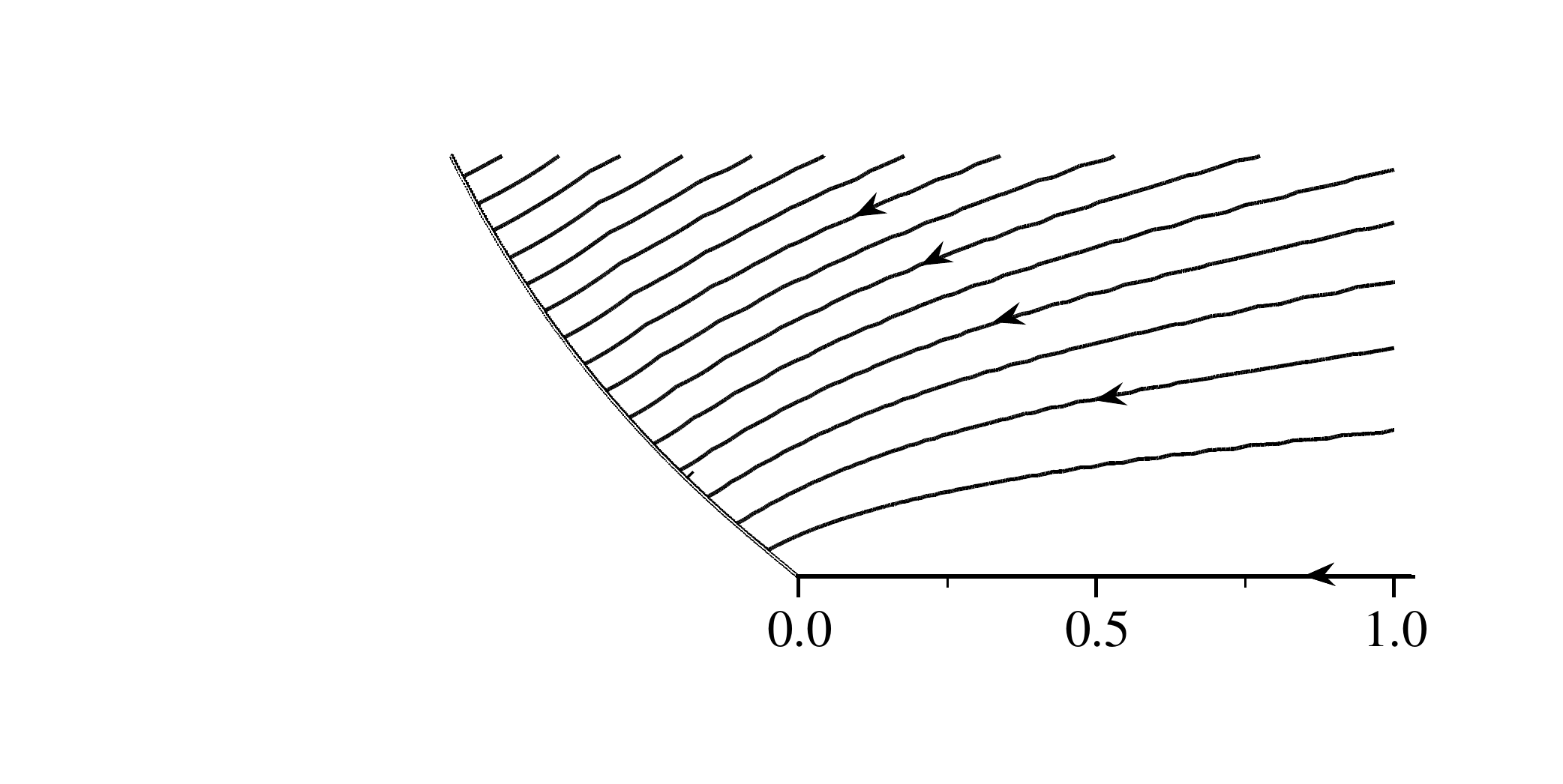}
\caption{Snapshot at $t=1.5$ of the streamlines in a spreading liquid drop. The solid surface
corresponds to $\psi=0$. $A$ is computed to be $1.6025$. Streamlines in all plots are in increments
of $\psi=-0.05$.  Top: plot showing the computed velocity field in both the inner and outer
regions. Bottom: plot showing the actual streamlines obtained after superimposing the eigensolution
with the calculated solution in the inner region.} \label{F:135transstreamlines_unsteady}
\end{figure}

Finally, we show that our method is equally applicable to
time-dependent flows. As an illustration, we consider the spreading
of a two-dimensional liquid drop over the solid surface, with the
axis of symmetry at $x=0$.  The drop is driven from its initially
cylindrical shape by gravity. In
Fig.~\ref{F:135transstreamlines_unsteady}, we show a snapshot of the
streamlines near the contact line in a frame moving with the contact
line. The free surface is in a state of evolution and hence no
longer represents a streamline. Again, a comparison of the two plots
in the figure show that the position of the transition line does not
effect the overall flow.

\section{\label{sec:conclusion}Conclusion}

We have shown that straightforward application of a standard numerical method to a seemingly
ordinary problem can lead to completely unacceptable numerical artifacts, despite the fact that the
conventional preliminary asymptotic analysis of a possible source of difficulties (the corner
formed by smooth parts of the domain's boundary) does not flag up any concerns.

A surprising result from the present study is that errors in approximating the velocity field
manifest themselves as spurious behaviour in the pressure field. It is the presence of an
underlying eigensolution which creates these numerical artifacts in the pressure distribution,
despite the fact the eigensolution itself has a globally constant pressure.  Therefore, preliminary
analytic work should always include a search for possible eigensolutions, an analysis of their
properties and an attempt to numerically approximate them individually. Only using this procedure
have we been able to determine the origin, and hence solution to, the problems presented in
Section~\ref{num}.

We have shown that, if the eigensolution is not removed prior to computations, one invariably ends
up with huge pressure spikes whose position and magnitude are both mesh-dependent. The numerical
analysis suggests that the cause of this numerical instability is the errors in approximating the
velocity gradient, as in the eigensolution this gradient is singular at the corner. The `mechanism'
of this error generation and the subsequent instability require a thorough mathematical
investigation, which lies outside the goal of the present paper. Such an investigation would,
hopefully, point out other situations where standard numerical methods could run into trouble.

The developed method of removing numerical artifacts in the pressure distribution is not only
successful with respect to the model case of a steady two-dimensional Stokes flow in a corner
region, but, as is shown, it admits a straightforward generalization which makes it applicable to a
general case of unsteady free-boundary Navier-Stokes flow. In practical applications to problems of
dynamic wetting one often has a situation where in the process of computations the contact angle
varies in a wide range, from the angles where a standard numerical code produces no artifacts to
those where the pressure spikes and multivaluedness invariably appear. General-purpose numerical
algorithms should be developed so that the present method is turned off for acute angles, where the
pressure is naturally single-valued, and switched on for obtuse angles to suppress spurious
numerical behaviour.

The failure of standard numerical algorithms to approximate flow in a corner is not limited to the
formulation considered in the present paper. For example, an alternative mathematical approach to
the moving contact-line problem \cite[see][]{dussan76,zhou90,somalinga00} is to prescribe the
velocity as an explicit function of distance from the corner, so that, in a frame moving with the
contact line, the velocity at the contact line is zero and it tends to the speed of the solid in
the far field. When the Dirichlet conditions of this type are applied instead of the Navier-slip
(i.e.\ Robin-type) condition that we have examined, one observes spurious numerical features,
though in a different range of the corner angles to those in the present paper. This case is
examined in a forthcoming paper.

\section*{Acknowledgements}
The authors kindly acknowledge the financial support of Kodak European Research and the EPSRC via a
Mathematics CASE award.

\bibliographystyle{jfm}
\bibliography{manuscript_arxiv}

\end{document}